\documentclass[fleqn,usenatbib,useAMS]{mnras}

\usepackage{graphicx}	
\usepackage{amsmath}	
\usepackage{amssymb}	
\usepackage{multicol}        
\usepackage{bm}		
\usepackage{pdflscape}	
\usepackage{longtable}  

\usepackage[T1]{fontenc}
\usepackage{ae,aecompl}
\usepackage{caption}

\usepackage{newtxtext,newtxmath}

\title[Stellar Populations of ETGs in CGs]{Stellar Population Properties of ETGs in Compact Groups of Galaxies}

\author[Moura et al.]{Tatiana C. Moura,$^{1}$
Reinaldo R. de Carvalho,$^{2}$
Sandro B. Rembold,$^{3}$
Marina Trevisan,$^{4}$
\newauthor
Andre L. B. Ribeiro,$^{5}$ 
Angeles P{\'e}rez-Villegas,$^{1}$
Francesco La Barbera$^{6}$,
Diego H. Stalder,$^{7}$
\newauthor
and Reinaldo R. Rosa$^{8}$
\\
$^{1}$ Universidade de S\~ao Paulo, IAG, Rua do Mat\~ao 1226, Cidade Universit\'aria, S\~ao Paulo 05508-900, Brazil\\
$^{2}$ NAT-Universidade Cruzeiro do Sul / Universidade Cidade de S\~ao Paulo\\
$^{3}$ Universidade Federal de Santa Maria 97105-900, Santa Maria-RS, Brazil\\
$^{4}$ Universidade Federal do Rio Grande do Sul, Departamento de Astronomia, CP 15051, Porto Alegre 91501-970, Brazil\\
$^{5}$ Universidade Estadual de Santa Cruz, Rodovia Jorge Amado km 16, Ilh\'eus 45662-000, Brazil\\
$^{6}$ INAF, Observatorio Astronomico di Capodimonte - I-80131 Napoli, Italy\\
$^{7}$ Universidad Nacional de Asunci\'on Facultad de Ingenier\'ia, Campus Universitario, Paraguay \\
$^{8}$ Instituto Nacional de Pesquisas Espaciais/MCTI -Av. dos Astronautas 1758
}

\pubyear{2019}

\begin{document}
\label{firstpage}
\pagerange{\pageref{firstpage}--\pageref{lastpage}}
\maketitle

\begin{abstract}

We present results on the study of the stellar population in Early-Type galaxies (ETGs) belonging to 151 Compact Groups (CGs).  We also selected a field sample composed of 846 ETGs to investigate
environmental effects on galaxy evolution. We find that the dependences of mean stellar ages, [Z/H] and [$\alpha$/Fe] on central stellar velocity dispersion are similar, regardless where the ETG resides, CGs or field.  When compared to the sample of centrals and satellites from the literature, we find that ETGs in GCs behave similarly to centrals, especially those embedded in low-mass haloes ($M_{h} < 10^ {12.5}M_{\odot}$). Except for the low-mass limit, where field galaxies present a Starforming signature, not seen in CGs, the
ionization agent of the gas in CG and field galaxies seem to be similar and due to hot, evolved low-mass stars. However, field ETGs present an excess of H$\alpha$ emission relative to ETGs in CGs. Additionally, we performed a dynamical analysis, which shows that CGs present a bimodality in the group velocity dispersion distribution - a high and low-$\sigma$ mode. Our results indicate that high-$\sigma$ groups have a smaller fraction of spirals, shorter crossing times, and a more luminous population of galaxies than the low $\sigma$ groups.
It is important to emphasize that our findings point to a small environmental impact on galaxies located in CGs. The only evidence we find is the change in gas content, suggesting environmentally-driven gas loss.

\end{abstract}





\section{Introduction}

Research in extragalactic astrophysics has made significant progress in the past fifteen years, mostly because of the large surveys that offered a deeper look into the Universe. Even with those undeniable advances, many open questions about the formation and evolution of galaxies still remain. We know that the environment plays a role in the evolution of galaxies, but the extension of that influence remains unclear, a good example of  which are the associations of galaxies known as Compact Groups (CGs). They show high spatial density, despite being composed by no more than ten galaxies, and present a moderate velocity dispersion ($\sim 200$ km/s), typical of galaxies in low density environments. Because of these properties, CGs are considered an ideal place for studying dynamical interactions and mergers. They offer all conditions required for a merge to happen -- high density and low relative velocities --, and early N-body simulations \citep{Barnes85} estimated that after $\sim$1 Gyr galaxies in CGs should merge into a ``fossil'' giant elliptical galaxy. Despite the fact that many groups show signs of mergers, the actual number of observed CGs is too high to fulfill such prediction. Later studies show that certain initial conditions \citep{DiMatteo2008} or the dark matter distribution \citep{Athanassoula97} could prolong the lifetime of CGs for at least $\sim$9 Gyr. Also, loose groups may be a source of replenishment for CGs \citep{Diaferio94, Governato96, Ribeiro98, Andernach2005, Mendel2011, Pompei2012}. The fact that isolated CGs show an excess of Early-Type Galaxies (ETGs) collaborates to such ``replenishment mode'' scenario \citep{Andernach2007}.

The type of gas ionization mechanism in CG galaxies may reflect these expected high merger rates and interactions. One way to trigger an AGN (Active Galactic Nucleus) is  galaxy-galaxy interaction that may feed a supermassive black hole (SMBH) in the center of the galaxy. These interactions can also induce star formation if enough gas is involved in the process. In fact, ionization by AGN is conspicous in CGs, both in form of LINERs and low luminosity, high-ionization nuclear activity \citep{Coziol1998, Coziol2000, Martinez2008, Gallagher2008, Martinez2010}. The deficiency of gas in these groups can explain the type of activity and the absence of new star formation episodes. \cite{Verdes-Montenegro2001} conclude, analyzing 72 systems defined by \cite{Hickson82}, that CGs are $\sim 40 \%$ depleted in HI from the expected value based on the optical luminosity and morphology of the member galaxies. 

CGs are richer in elliptical galaxies than the field \citep{Lee2004,Deng2008}. Galaxies in CGs also tend to be older than galaxies in the field \citep{Proctor2004, deLaRosa07, Plauchu2012} but have similar ages when compared to clusters \citep{Proctor2004}. This is often interpreted as an indication that CGs speed up the evolution of galaxies from star forming to quiescent \citep{Tzanavaris2010, Walker2010, Coenda2012}. Some authors even find evidence of truncation in the star formation for galaxies in CGs \citep{deLaRosa07}. All these observations reinforce the scenario where CGs are gas-poor systems. 

In this work, we investigate the stellar population properties of ETGs in CGs and their relation to the dynamics of these groups and gas ionization source present in these galaxies. For a better understanding of the effects caused by the CG environment, we compare our results to those obtained for a sample of ETGs in the field (low density)
and to those determined for a sample of central galaxies studied by  \cite{SPIDER10}.

This paper is organized as follows. In Section \ref{sec:sample}, we describe the samples in different environments and how we discriminate galaxies of different morphologies. Section \ref{sec:stellar_pop} presents the methods applied to estimate the stellar population parameters. An analysis of the gas ionization agent in our sample is discussed in  Section \ref{sec:activity}, followed by a dynamical analysis of our sample of CGs in Section \ref{sec:dynamic}. We discuss our results and present a summary in Section \ref{sec:discussion}.

\section{Sample and Data}\label{sec:sample}

The focus of this study is the ETGs belonging to CGs. For the definition of the sample,  we choose the extensive catalogue of bright galaxies in CGs defined by \cite{McCon09}. After the selection, we performed a visual classification to select the elliptical galaxies belonging to CGs. For this, we mimic the same scheme applied in the second version of ``The Galaxy Zoo Project'' \citep{Zoo2}. We also defined a control sample, constituted of ETGs in the low-density environment (field), as we discuss below.

\subsection{The Compact Group Sample} \label{subsec:cg}
\smallskip

Our sample of CGs was extracted from the ``Catalogue A'' compiled by \cite{McCon09}. This catalogue includes objects classified as galaxies and with magnitude in the $r$ band brighter than  $m_{r} = 18$ in the database of the sixth release of the Sloan Sky Digital Survey (SDSS DR6). Compact groups were identified using the three well-known photometric criteria determined by \cite{Hickson82}. The first criterion, \emph{population}, specifies that CGs must be composed by at least four members in the magnitude range [$m_{1}$,\,$m_{1}+3$], where $m_{1}$ is the magnitude of the brightest group member. The \emph{compactness} criterion states that the group mean surface brightness, $\langle\mu_{e}\rangle$, must be brighter than 26 mag/arcsec$^{2}$ in the $r$-band. The last one, the \emph{isolation}, establishes that no objects in the same magnitude range as the CG member galaxies are present in a ring of angular size $\theta_{G} \geq 3\theta_ {N}$, where $\theta_{N}$ is the smallest concentric circle encompassing the centre of the galaxies defining the group. Taking all these criteria into account, the catalogue ends up with 2297 CGs (9713 galaxies) covering the magnitude range $14.4 \leq r \leq 18.0$ and redshift $z\leq 0.2$. By visual inspection of the objects in Catalogue A, the authors eliminate the contamination by photometric errors from the SDSS algorithm, such as misclassified objects, satellite tracers and saturated objects. However, when considering the spectroscopic data, the authors estimate that 55\% of the CGs present interlopers.

\begin{figure}
\includegraphics[width=0.49\textwidth]{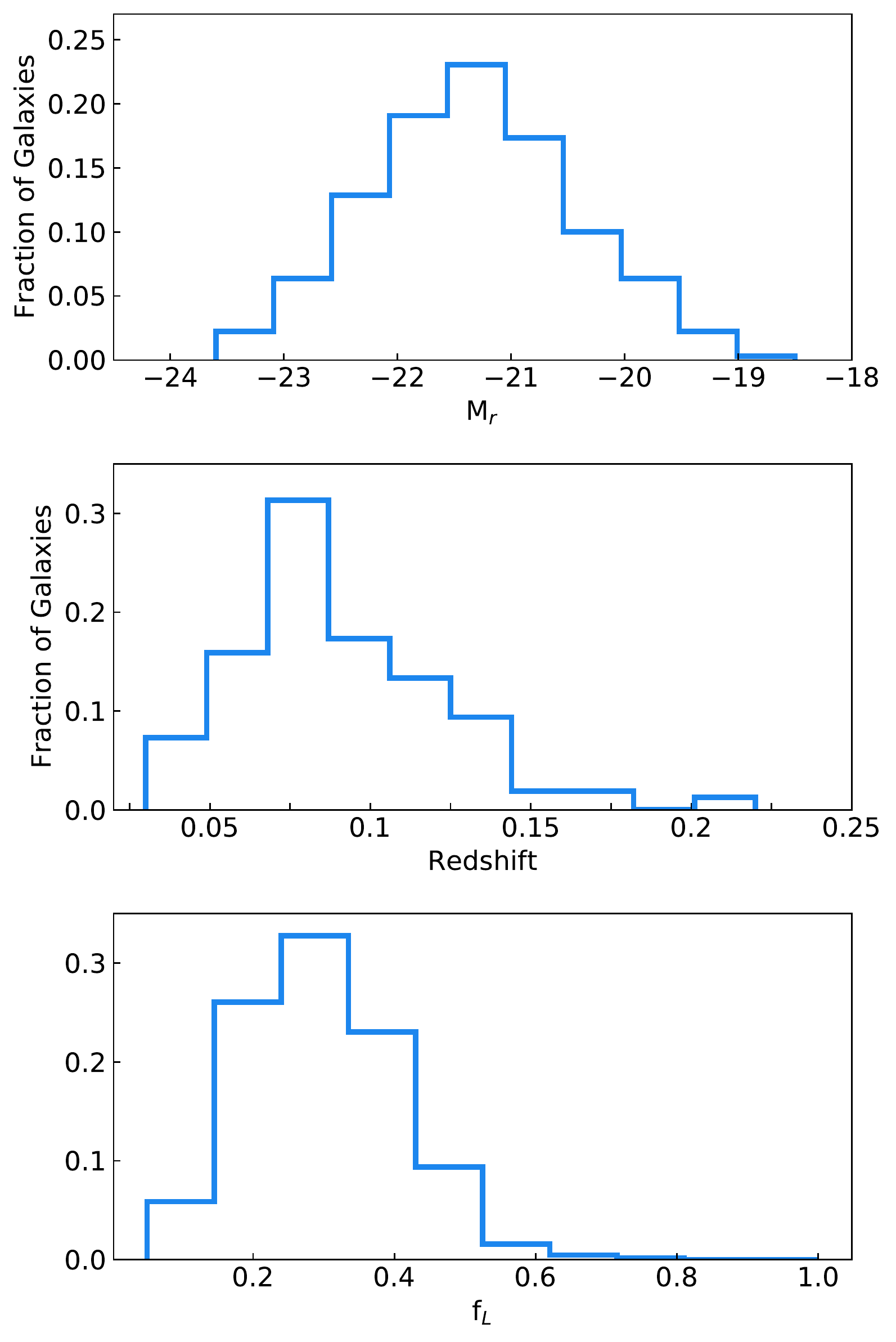}
\caption{Absolute magnitude ($M_{r}$), redshift and fraction of light ($f_{l}$) distributions for the 
629 galaxies that composed our sample of GCs. The parameter $f_{l}$ is defined as  the fraction of light captured by the SDSS fiber aperture. \label{fig:cg_dist}}
\end{figure}

To increase the spectroscopic data available, we have searched for the galaxies from Catalog A in the database of the twelfth release of SDSS (DR12) \citep{SDSSDR12}. We found that the spectra of $\sim 53\%$ (5353 galaxies) of the objects in Catalog A are available in DR12. From that initial sample, we selected only groups with at least 4 members with redshifts satisfying the concordant redshift criterion ($\Delta cz \leq 1000$ km/s) as in \cite{Hickson92}.We do not include a colour criterion given the well-know degeneracy with the age and metallicity which are parameters that we are interested in investigating. Our final sample of CGs is composed by 629 galaxies distributed in 151 GCs. Some galaxy properties such as absolute magnitude $M_{r}$, redshift and fraction of light captured by the optical fiber ($f_{L}$) are presented in Figure \ref{fig:cg_dist}. Table \ref{tab:Nz} summarizes, for the whole sample, the number of groups $N_{groups}$, with $N_{z}$ members with redshift available and the total number of members, $N_{members}$.

\begin{table}
  \centering
  \caption{Summary of the number of group members with spectra available in SDSS DR12. We list the number of groups, $N_{groups}$, with $N_{z}$ members with redshifts available and the total number of members, $N_{members}$.}
  \label{tab:Nz}
  \begin{tabular}{ccc}
  \hline\hline
  $N_{z}$ & $N_{members}$ & $N_{groups}$ \\
  \hline
   & 4 & 81 \\
   & 5 & 30 \\
 4 & 6 & 10 \\
   & 7 & 5\\
   & 8 & 3 \\
\hline
   & 5 & 13 \\
 5 & 6 & 6 \\
   & 9 & 1 \\
\hline
 6 & 6 & 1 \\
\hline
 7 & 8 & 1 \\
\hline
  \end{tabular}
\end{table}

\subsection{Morphology for the CG Sample}

For the morphological selection, we apply the same methodology used in The Galaxy Zoo Project \citep{Zoo1, Zoo2}. This project, after more than a decade of existence, has produced four catalogues of galaxy morphological classifications. We searched, in the first and second versions of the catalogue, the morphological classification for the 629 galaxies that compose our sample. From the first catalogue (hereafter ``Zoo1''), around 70\% (441) of the galaxies in our sample have morphology determined, although 309 out of those 441 systems are listed as ``Unknown''. 
For the second version of the catalogue (hereafter ``Zoo 2''), more restrictions were applied for the selection of the objects in terms of size (petroR90$_r >$ 3 arcsec, where petroR90$_r$ is the parameter that measures the radius containing 90\% of the Petrosian flux in the $r$ band), brightness (Petrosian half-light magnitude $m_{r} <$17.0) and redshift (0.0005 $< z <$ 0.25). Since Zoo 2 presents morphological classifications for only the brighter galaxies in SDSS DR7, the number of objects from our sample found in the catalogue was correspondingly low (only 331), with most of them classified as ellipticals ($\sim 54\%$).

\begin{figure}
 \centering
 \includegraphics[width=0.48\textwidth]{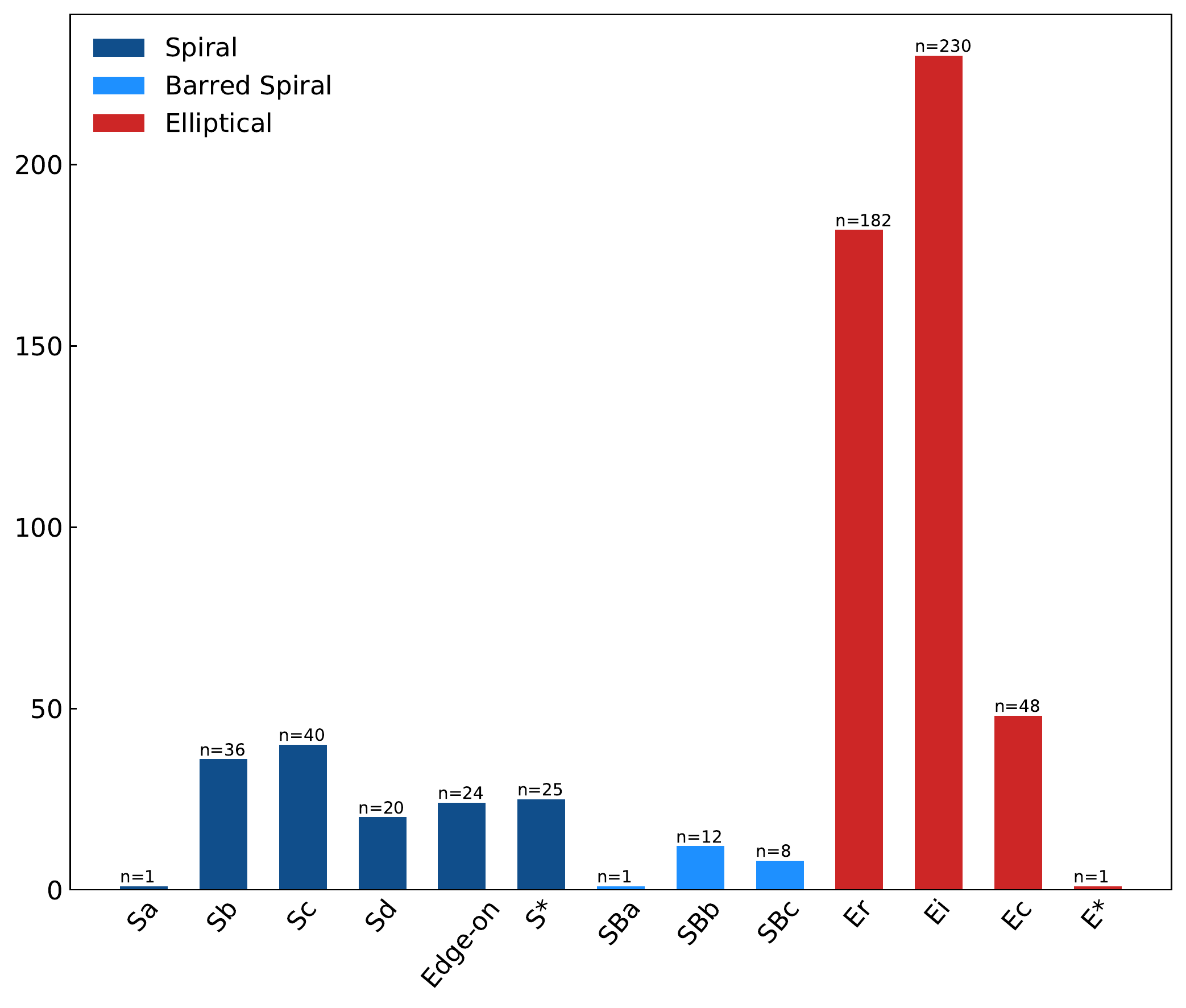}
 \caption{Distribution of galaxies according to the morphological classification performed by five voters for the galaxies in our CG sample.}
 \label{fig:MyZoo_class}
\end{figure}

The lack of morphological classifications for low-brightness galaxies, led us to reproduce the same form and use the decision tree from Zoo 2 and apply to the galaxies of our sample. A total of five persons responded to the questionnaire, and the most voted answers determined the class attributed for each object. Table \ref{tab:MyZoo} presents a list of the classification categories and Figure \ref{fig:MyZoo_class} shows a summary for all the 629 galaxies from our sample. A brief explanation of each category from Zoo 2 is given in Table \ref{tab:zooIIclass} in the Appendix. Our rating leads to a sample composed mostly by elliptical galaxies ($\sim 84\%$). For one elliptical galaxy, the voters did not specify the shape, and for 25 spirals there is no information about the presence of the bulge. In this last case, the galaxy is listed as ``S''.

\begin{table}
\centering
\caption{The morphological classification of the galaxies belonging to our MC09 sample. The table 
contains 629 lines but only the first 10 objects are reproduced here. We adopted the same nomenclature from The Galaxy Zoo 2 Project, where the letter inside parenthesis means: (m) = merger, (l) = lens/arc, (r) = rings, d = disturbed, i = irregular, (o) = others and (d) = dust lane. The ``GroupID'' is the number of the group given 
in \citet{McCon09}, ``GalID'' is the position of the galaxy  in order of brightness in the group 
and ``ObjID'' is the identification of the object in SDSS DR12 database. The full table is available online.}
\label{tab:MyZoo}
\begin{tabular}{lccc}
\hline\hline
GroupID & GalID & ObjID & Class \\
\hline
  42 & 1 & 1237661137960632449 & Er \\
  42 & 3 & 1237661137960632448 & Ei \\
  42 & 4 & 1237661137960632447 & Ei \\
  42 & 2 & 1237661137960632446 & Ei \\
  46 & 4 & 1237654390032629949 & Er \\
  46 & 3 & 1237654390032629946 & Ei \\
  46 & 2 & 1237654390032629944 & Ei \\
  46 & 1 & 1237654390032629943 & Er \\
  70 & 4 & 1237662224058024106 & Ei \\
  70 & 2 & 1237662224058024105 & Ei(m)\\
  \hline
\end{tabular}
\end{table}

\begin{figure}
 \centering
 \includegraphics[width=0.4\textwidth]{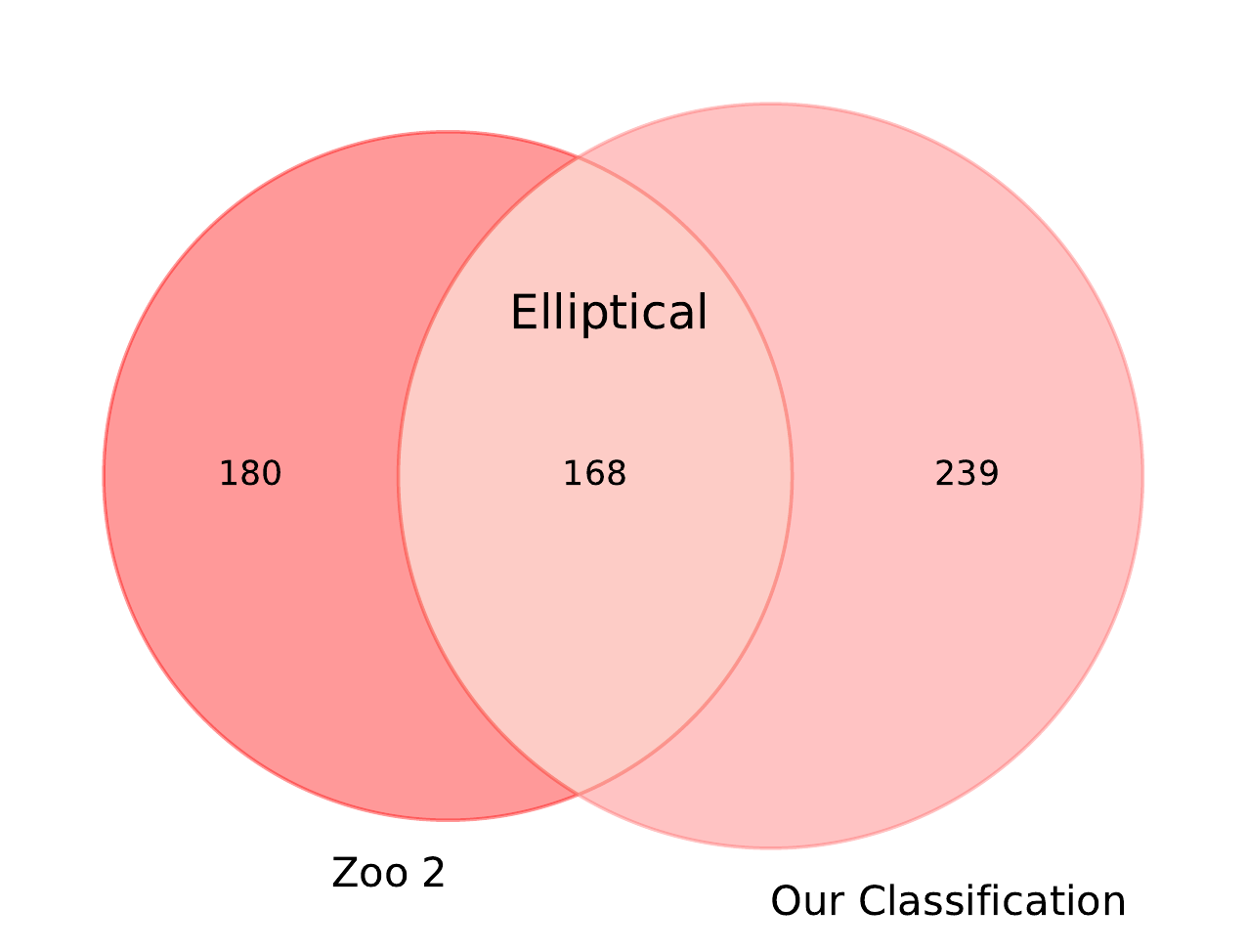} \\ \includegraphics[width=0.4\textwidth]{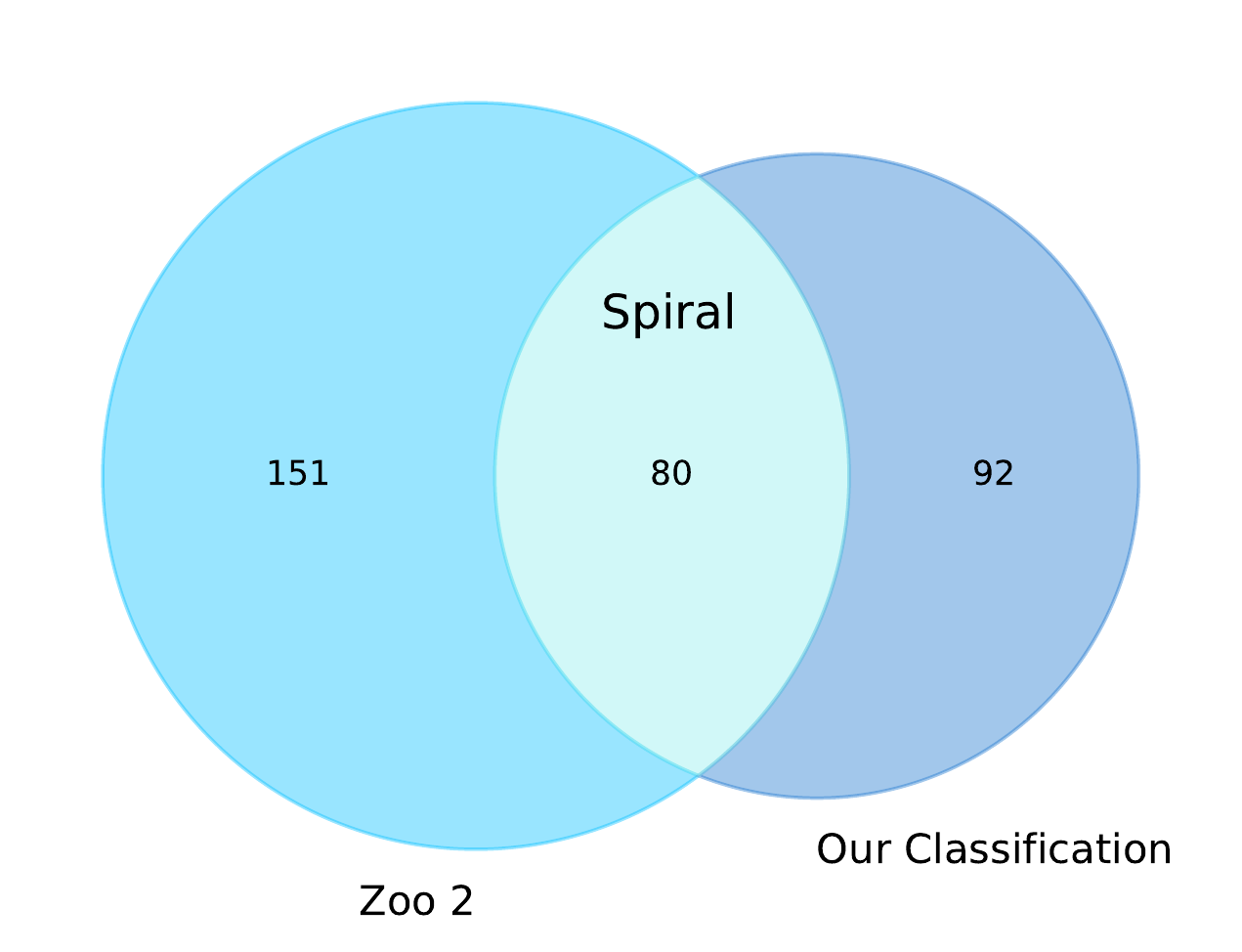}
 \caption{Comparison between the classifications of 331 galaxies of our sample that also belong to the Galaxy Zoo 2 catalogue. We apply the same methodology of the second version of the project to perform our morphological classifications.}
 \label{fig:Zoo2_Venn}
\end{figure}

To estimate how reliable our classification is, we compare our result for those 331 galaxies in common with the Zoo 2 database, as shown in Figure \ref{fig:Zoo2_Venn}. According to Zoo 2, 180 galaxies among those 331 systems are ellipticals. Our classification is in agreement for 168 from those 180 elliptical galaxies (superior panel of Figure \ref{fig:Zoo2_Venn}). Considering the completeness as the fraction of galaxies with the same class in both classification schemes,  we estimated a completeness of $\sim93\%$ . For estimating the contamination in our classification, we count the number of spirals given in Zoo 2 catalogue  (151 galaxies) which are classified as ellipticals in our experiment (71 galaxies), as shown in the inferior panel in Figure \ref{fig:Zoo2_Venn}. This leads to contamination rate of  $\sim 51\%$. This high contamination could be the reflex of the low number of voters in our classification compared with the thousands of voters from Zoo 2; however, all voters in our experiment are experienced astronomers. Our final sample of ellipticals in CGs are those 461 galaxies classified as such by our voters.

\subsubsection{Field ETGs Sample} \label{subsec:field}

\begin{figure}
 \centering
 \includegraphics[width=0.48\textwidth]{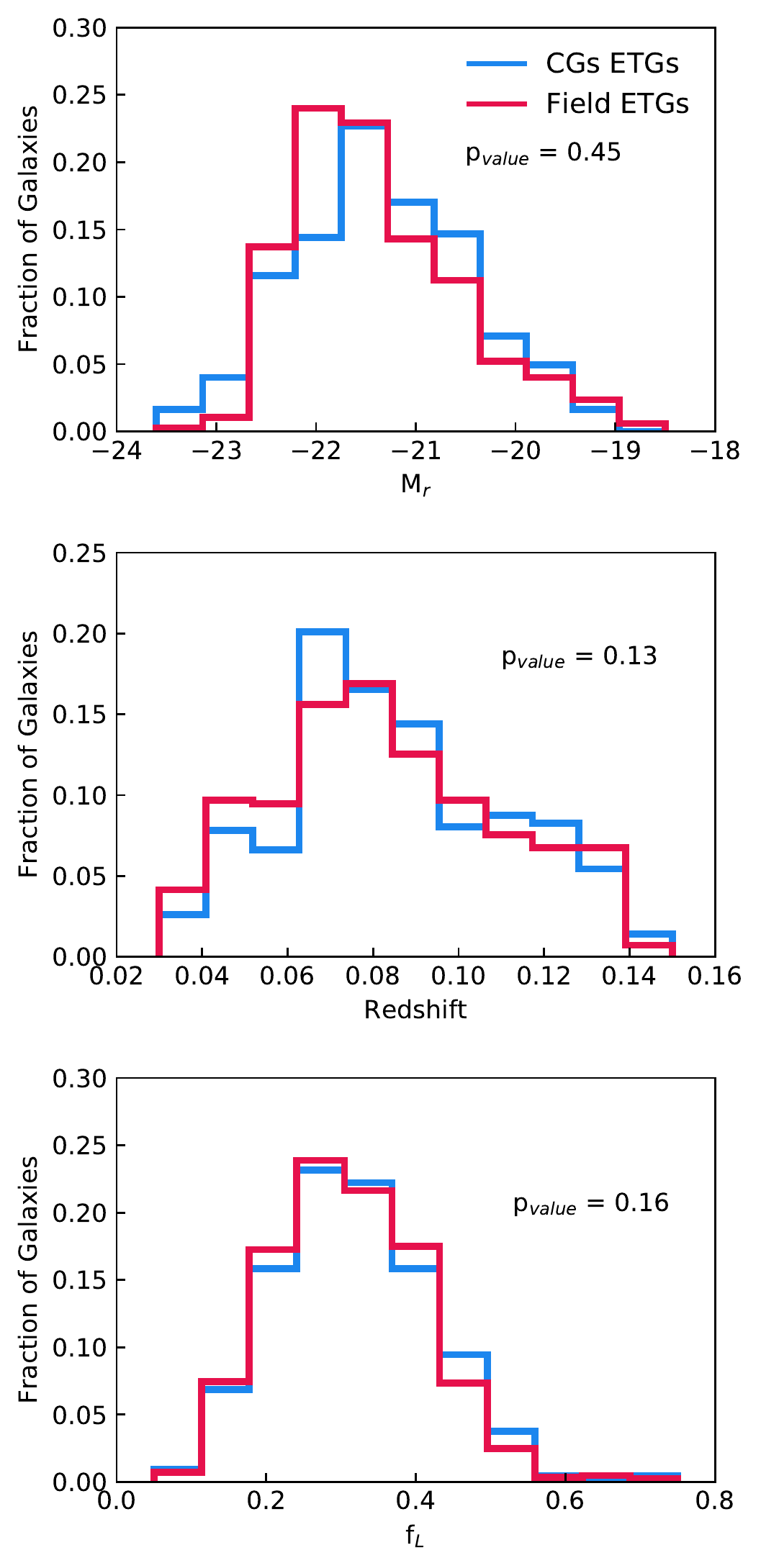}
 \caption{Comparison between the distributions of redshift, $k$-corrected absolute magnitude in the $r$ band ($M_{r}$) and the fraction of light ($f_{L}$) for the field and CGS samples. The $p$-value was estimated through a permutation test.}
 \label{fig:field_dist}
\end{figure}

In order to perform a consistent analysis of the effects of the environment on galaxies in GCs, we have selected a control sample of elliptical
galaxies in the low density environment of the field. This allow a rich comparison between the properties
of these galaxies in different environments. For the field galaxy sample we selected  only galaxies that are more distant than 10 $R_{vir}$ from all groups with halo masses greater than 10$^{13}$ $M_{\odot}$, 
following the approach described in \cite{Trevisan2017}. We used the updated version of the group catalog 
compiled by \cite{Yang07}. This updated catalogue contains 473\,482 groups drawn from a sample of galaxies 
mostly from the SDSS-DR7 \citep{Abazajian2009}. Although the catalogue by \cite{Yang07} contains objects up to $z \leq 0.2$, we cut our sample in z = 0.14, since the catalogue is complete for groups with halo masses $\gtrsim 10^{13}\,$M$_{\odot}$ below this redshift (see Figs. 5 and 6 in \citealp{Yang07}). To be consistent, we applied the same redshift cut to the ETGs in CGs, leading to a sample with 423 CG galaxies used for the matching procedure. The initial field sample is then composed by  $130\,767$ galaxies. We then selected only the galaxies that are classified as elliptical according to the Galaxy Zoo 2, reducing the sample to $17\,499$ galaxies. To assure
that the galaxy is not close to a group or cluster outside the borders of the SDSS, we require that at 
least 95 per cent of the region within 500 kpc from the galaxy lies within the SDSS coverage area.
For this purpose, we adopted the SDSS-DR7 spectroscopic angular selection function mask provided by the 
NYU Value-Added Galaxy Catalog team \citep{Blanton2005} and assembled with the package MANGLE 2.1
\citep{Hamilton2004,Swanson2008}. We excluded $3\,179$ galaxies that do not satisfy this criterion. Finally, we extracted from this sample of $14\,320$ objects a control sample of 846 galaxies (twice the size of the GC sample) with similar stellar masses, at similar redshifts and with similar fraction of light within the SDSS fiber as the CG sample by applying the Propensity Score Matching (PSM) technique \citep{PSM}. For the PSM, we used the MatchIt package \citep{Ho2011} written in R \citep{R2015}. This technique allows us to select from the sample of field galaxies a control sample in which the distribution of observed properties is as similar as possible to that of the CG galaxies. We adopted the Mahalanobis metric \citep{Mahalanobis36} and the nearest-neighbour method to perform the matching. In Figure \ref{fig:field_dist} we compared the distribution in $M_{r}$, redshift and $f_{L}$ for both samples used in this work. We also executed a permutation test in order to check if the distributions are indeed similar. The p-value for the distribution of absolute magnitude (p = 0.45), redshift (p = 0.13) and the fraction of light within the SDSS fiber (p = 0.16) allows to reject the null hypothesis and consider that the samples came from the same parent population.

\section{Stellar Population Parameters} \label{sec:stellar_pop}

A way to characterize a stellar population is determining quantities like mean stellar age, metallicity, [Z/H], and alpha enhancement, [$\alpha$/Fe]. There are two widely used  techniques to recover those parameters: spectral fitting and spectral index analysis. In the following, we describe how we combined both techniques to estimate all relevant stellar population parameters for our samples of ETGs. 

For better results, we limited our sample to galaxies which spectra provide a signal to noise ratio of $S/N \geq 15$ and velocity dispersion between $70\leq \sigma_{0}\leq 420$ km/s. This final cut leads to a sample of 303 ETGs in CGs and 697 in the field. In the next section, we compare the results for these samples.

\subsection{Spectral Fitting}\label{subsec:spectrafitt}

For our spectral-fitting methodology, we consider that a galaxy spectrum can be represented as a linear combination of a set of Single Stellar Populations (SSPs). We use a set of 108 SSPs extracted from the extended MILES (MIUSCAT) library \citep{Vaz10} covering stellar populations with 27 ages between 0.5 and 17.78 Gyr and four metallicities -- $[M/H] = \{-0.71, -0.40, 0, 0.22\}$. These models use the Padova isochrones and a Kroupa Universal Initial Mass Function. The SSPs cover the wavelength interval from 3465 to 9469 \AA\, with a spectral resolution of $\sim$2.51 \AA\, (FWHM). This is the same set used in the SPIDER Project \citep{SPIDER10}.

The full spectral fitting is performed with the STARLIGHT code \citep{Starlight1, Starlight2, Starlight3}. Before running the code, the observed spectra are corrected for foreground Galactic extinction and shifted to the rest frame. As for the models, we degraded the spectra to the mean resolution of SDSS (3 \AA{}). We performed the fitting in the wavelength interval of 4000 - 5700 \AA{}, which excludes the blue regions where the abundance ratio of non-solar elements could lead to a bias when we use nearly solar SSPs (Miles). This interval also excludes the regions with presence of molecular bands such as TiO that cannot be well-fitted with solar-scale models and a Kroupa IMF \citep{SPIDER8}. As for the extinction law, we select the more appropriate for elliptical galaxies, given by \cite{Cardelli}. The program output gives the ``population vector'' ($x_{j}$), which is the fraction of the total light that each SSP contributes to the fitting. From the spectral fitting, we derive the mean stellar age as a function of  $x_{j}$ through

\begin{equation}
 <\log t_{\ast}> = \sum\limits_{j=1}^{N_{\ast}} x_{j} \log t_{j} \Big/  \sum\limits_{j=1}^{N_{\ast}} x_{j},
 \label{eq1}
\end{equation}

\noindent where $t_{j}$ is the age of the \textit{j}th SSP.

%
\subsection{Spectral Index}\label{subsec:specindex}

To complete the set of stellar population parameters, we use the spectral index technique to estimate the stellar metallicities and [$\alpha$/Fe]. We measure the line strengths of the lines Fe5270, Fe5335, Fe4383 and  Mg$_{b}$5177 using the code  {\fontfamily{pcr}\selectfont indexf} (\cite{Cardiel}). From the iron indices, we estimated the $\langle Fe_{3}\rangle$\footnote{$\langle Fe_{3}\rangle \equiv \frac{1}{3}(Fe5270+Fe5335+Fe4383)$}, an index sensible to [Z/H]. We also use the $[MgFe]'$ index as defined by \cite{TMB03} ($[MgFe]' = \{Mgb[0.72(Fe5270)+0.28(Fe5335)]\}^{1/2}$) for estimation of the [$\alpha$/Fe] parameter.

To remove the effect of the velocity dispersion we applied the broadening correction to the spectral index as defined in \cite{deLaRosa07}. The correction is the ratio between the index measured with a given velocity dispersion and the one measured in the rest frame ($\sigma$ = 0). The ratios are determined using the indices measured from the spectra produced using the model from \cite{Vaz10} customised for a set of velocity dispersions (50 - 350 km/s). Our correction is in excellent agreement with those applied by \cite{deLaRosa07}, which use the models from \cite{Vaz99}.


\subsection{Alpha enhancement}\label{subsec:alpha}

In the study of stellar populations, the [$\alpha$/Fe] parameter holds valuable information about the formation and 
evolution of the galaxy. The Fe and $\alpha$-element abundances relevant for the [$\alpha$/Fe] parameter are products of the final stages of the evolution of massive stars, where Fe comes mainly from the type Ia supernova while  $\alpha$-elements are produced by core-collapse Supernovae explosions. Stellar populations are formed from the gas present in Intergalactic Medium (IGM) and this medium is enriched in metals by supernova explosions or stellar winds. In this sense, by recovering their relative abundances of stellar populations we are also tracing their formation history.

The estimation of [$\alpha/Fe$] was made through a solar scaled proxy. The proxy is defined as the difference between two independent metallicities: $[Z_{Mgb}/Z_{Fe}] \equiv [Z/H]_{Mgb}-[Z/H]_{Fe}$. The metallicities are calculated fixing the age coming from the spectral fitting and by a polynomial fit with the metallicities and the indices $Fe_{3}$ and $Mg_b$ from the MILES models \citep{Vaz10}. Finally, for the [$\alpha$/Fe] we use the relation defined in \cite{SPIDER8}: $[\alpha/Fe]=0.55[Z_{Mgb}/Z_{Fe}]$.

 \subsection{Hybrid Method}\label{subsec:hybridmet}
 
 The result from the hybrid method is a combination of the age obtained from the spectral fitting and the parameters $[Z/H]$ and $[\alpha/Fe]$ estimated from the spectral index method. The [Z/H] value is calculated using the approach described for the $Z_{Fe}$ and $Z_{Mg}$, but now we perform a polynomial fit using the metallicity and $[MgFe]'$ index from the MILES models. The [$\alpha$/Fe] parameter is estimated as described in Section \ref{subsec:alpha}. 
 
\begin{figure*}
 \centering
 \includegraphics[width=1\textwidth]{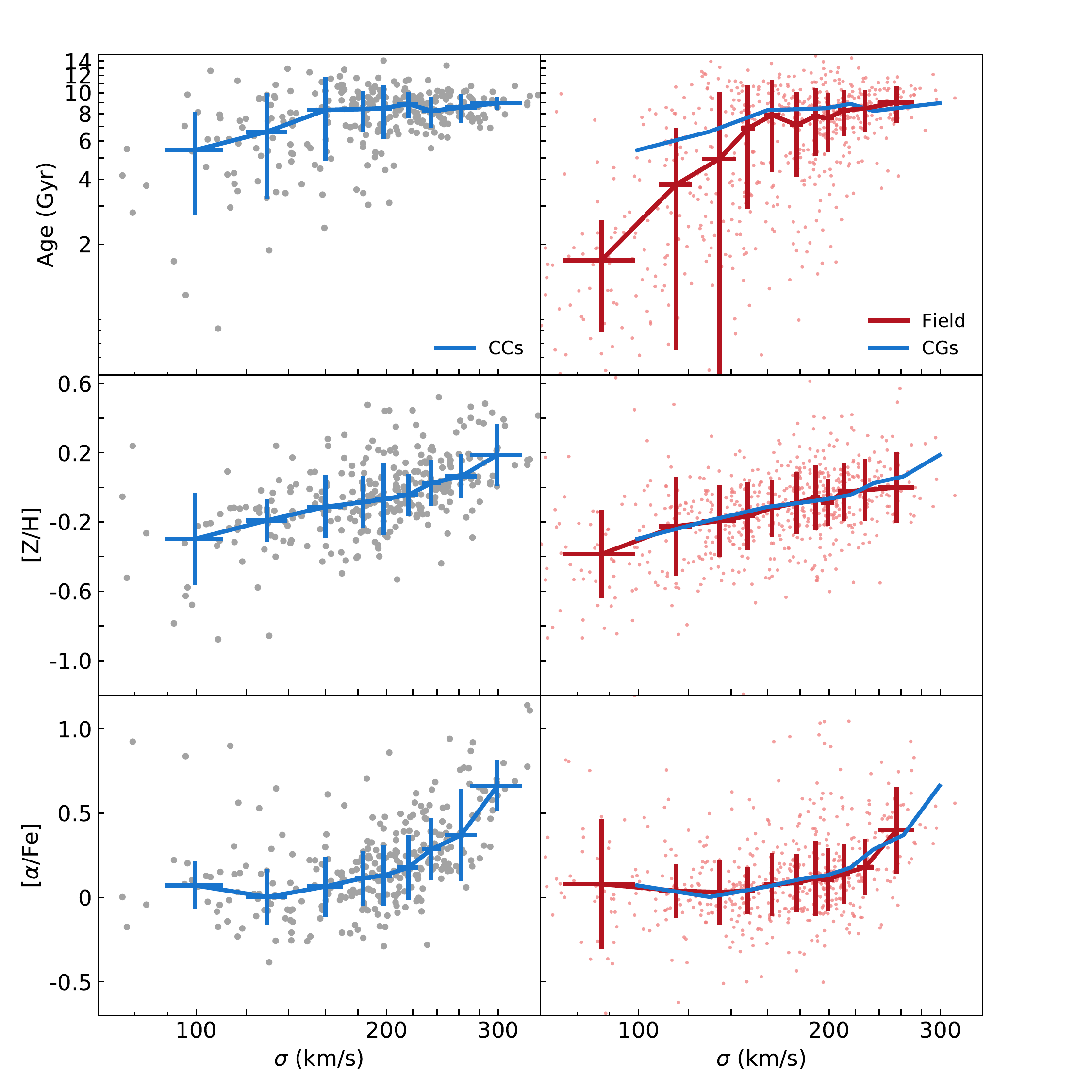}
 \caption{Stellar population parameters obtained from the application of the hybrid method. On the left panels, we show the mean stellar age, [Z/H] and [$\alpha$/Fe] trends for the 302 ETGs in CGs. In the right panels we compare the trends for the field sample (682 galaxies -- in red) and the CGs sample. The age is estimated from the spectral fitting and the [Z/H] and [$\alpha$/Fe] are calculated using a proxy. The [Z/H] proxy is given from the polynomial fitting of the metallicity and values of the $[MgFe]'$ from the MILES models \citep{Vaz10}. The [$\alpha$/Fe] is calculated using the relation from \citet{SPIDER8}.}
 \label{fig:hybrid_CGxField}
 \end{figure*}

 In Figure \ref{fig:hybrid_CGxField}, we present the stellar population parameters  as a function of the central velocity dispersion for ETGs in CGs (blue dots) and field (red dots). Since the central velocity dispersion depends on the distance and the aperture size of the optical fibre, it is necessary to apply an aperture correction. We use the correction given as a power law by \cite{Jorgensen95}, $\log (\sigma_{ap}/ \sigma_{n}) = -0.04  \log(r_{ap}/r_{n})$, where $\sigma_{ap}$ is the velocity dispersion from SDSS DR12 measured through an aperture $r_{ap} = 1.5'$ for the spectrograph used in the Legacy SDSS program or $r_{ap} = 1.0'$ for the objects observed in the BOSS program. We set $r_{n} = 1/8 r_{e}$, where $r_{e}$ is the effective radius; in this case, we use the de Vaucouleurs radius given by the deVRad$_r$ parameter from the DR12 SDSS database. A wrong sky subtraction or weak spectral lines could compromise the index measure providing unrealistic [$\alpha/Fe$] in the final application of the hybrid method. Because of such errors, a total of 30 galaxies from the field sample are not included in our results. From the CGs sample, only one ETG was excluded for the same reason.

Our result show that the stellar populations present in ETGs belonging to CGs behave similarly as the ETGs in the field. The stellar population parameters from both samples increases towards systems with higher central velocity dispersion. Once the velocity dispersion is an indirect measure of the dynamical mass of the system, our results show that massive galaxies are older, more metal-rich and with higher $[\alpha/Fe]$ than the less massive (low-$\sigma$) ETGs. The only noticeable difference is for the age in the low-$\sigma$ regime ($\sigma \leq \sim 130$ km/s), where the ETGs in CGs seems to be older ($\sim 2$ Gyr) than the ones in the field.  

 \section{Activity Analysis}\label{sec:activity}
 
Following the purpose of establishing differences between ETGs in the environments of CGs and field, we also analysed the type of ionization sources responsible for the emission lines in our sample. For such, we measured relevant emission lines fluxes and equivalent widths after stacking our individual spectra to increase the signal-to-noise ratio, and used diagnostic diagrams to perform the classification, as we describe below.
 
\subsection{Stacked Spectra} \label{subsec:stacked}

Optical diagnostic diagrams rely on emission line ratios, which are prone to significant uncertainties if the individual line fluxes are not well constrained. Those diagrams also require spectra with a high signal-to-noise ratio to minimise errors in the calculation of line ratios. It could be even more challenging if we are dealing with galaxies presenting weak emission lines, such as ellipticals. The spectra of our ETGs samples present typical S/N values in the order of $\sim$20; because of this, for the gas ionization source analysis, we have used stacked spectra. Stacking spectra allows for an increase in the S/N ratio, which leads to a more reliable result. The stacked spectra were produced by median-combining the individual normalized spectra in bins of velocity dispersion between $\sigma = 70-300$ km/s. The bin widths were determined in a way that each bin has a certain minimal number of galaxies. For the sample of ETGs in CGs we defined a minimum of $N_{bin}\geq 20$ galaxies per bin, and for the field sample, we used $N_{bin} \geq 60$. In this way, the width of the bins varies from 10 to 50 km/s. Our analysis is based on eleven spectra for the CGs sample and eight from the field.

 \subsection{Diagnostic Diagrams}
 
\begin{figure*}
\centering
\includegraphics[width=.54\textwidth]{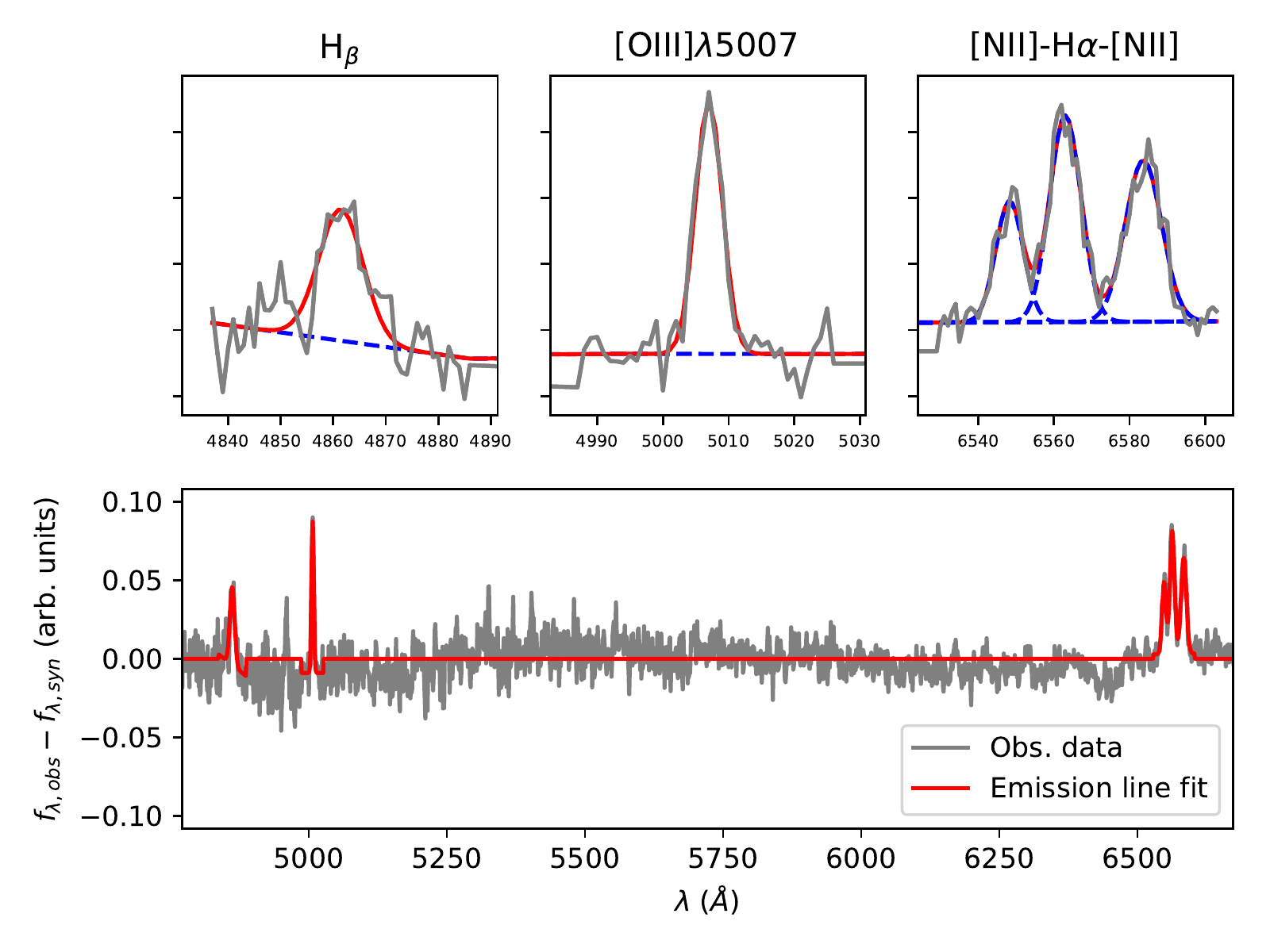}
\caption{An example of the Gaussian fitting used to measure the fluxes of the emission lines in the stack spectra. The top panels show the individual fitting of the H$\beta$ and [OIII]$\lambda$5007 lines and the simultaneous fitting for the triplet [NII]$\lambda$6548-H$\alpha$-[NII]. In the bottom panel, we exhibit the full spectra and the detected emission lines.}
\label{fig:ex_pylineflux}
\end{figure*}

We have used the diagnostic diagrams defined by \citet{BPT} and \citet{Cid Fernandes2011} to classify the type of activity in our ETG samples. In order to correct
the emission features for stellar absorption, we have subtracted
from the stacked spectra the corresponding best-fit stellar population synthesis solution obtained with STARLIGHT. 
The fits described in Sect. \ref{subsec:spectrafitt} are not suitable for this purpose, since they do not extend to some important optical transitions above 6000\AA\,{} (e.g. the H$\alpha$ line). We
have therefore performed a new run of STARLIGHT similar to the previous one, but extending the fitting window to 6900\AA{}. After subtraction of the best-fitting models, the emission line fluxes of H$\beta$, [OIII]$\lambda$5007, [NII]$\lambda$6548, H$\alpha$ and [NII]$\lambda$6584 were measured by Gaussian fitting the relevant spectral ranges in each stacked spectrum. This fit was done taking into account the uncertainties in all spectral pixels and the wavelength-dependent resolution of the SDSS spectra. The continuum around each line was allowed for a constant tilt, to account for possible mismatches between SSP models and the stacked spectra. The H$\beta$ and [OIII]$\lambda$5007 have been fitted individually, but a simultaneous fit was applied to the triplet [NII]$\lambda$6548-H$\alpha$-[NII]$\lambda$6584. Line equivalent widths have been obtained as the ratio between
the line fluxes and the median continuum at the central wavelength of the emission lines,
measured directly on the best-fit STARLIGHT spectrum. An example of the emission line fitting is shown in Figure \ref{fig:ex_pylineflux}. 

\begin{figure*}
\centering
\begin{tabular}{cc}
\includegraphics[width=.52\textwidth]{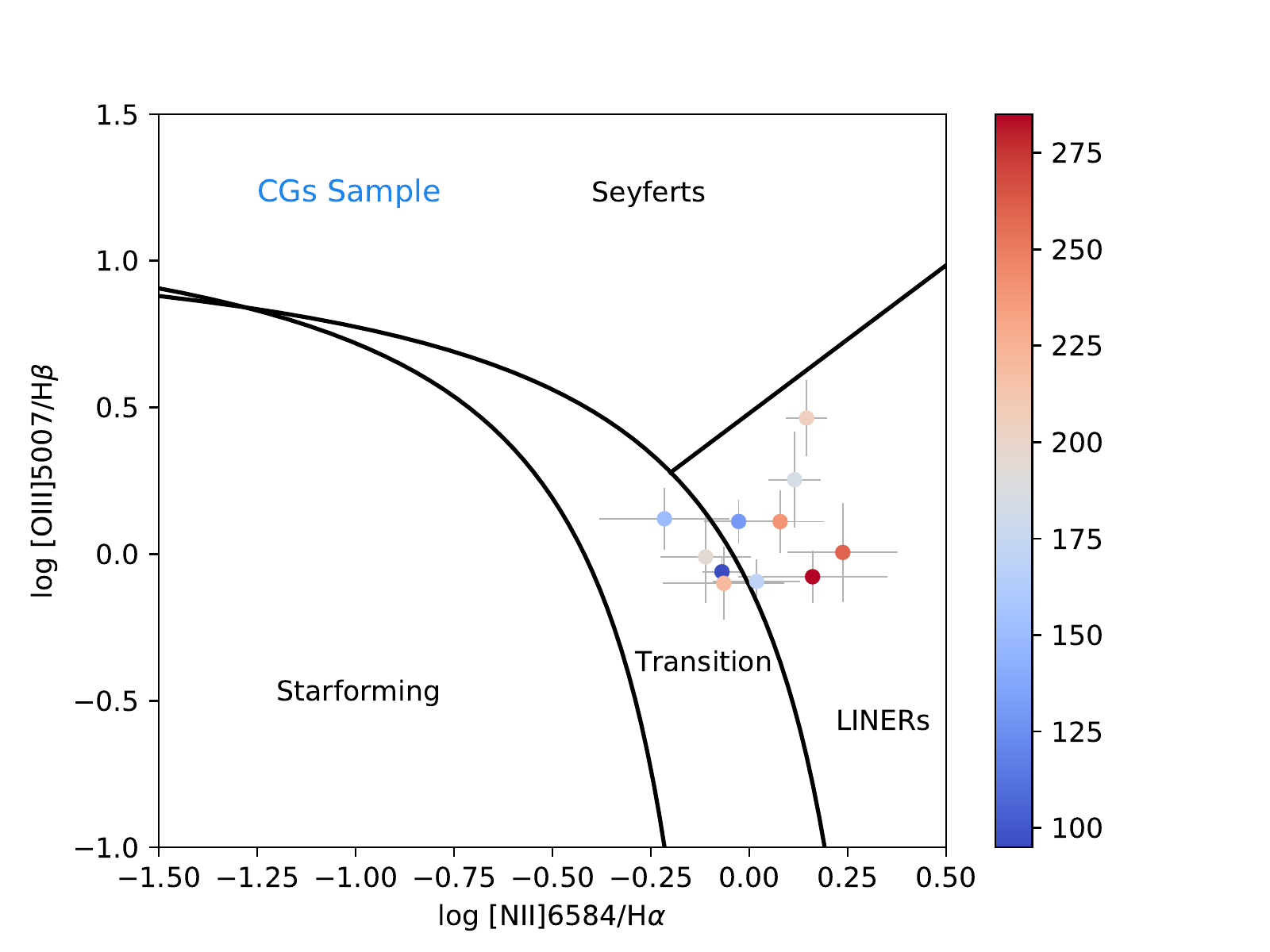} &
\includegraphics[width=.52\textwidth]{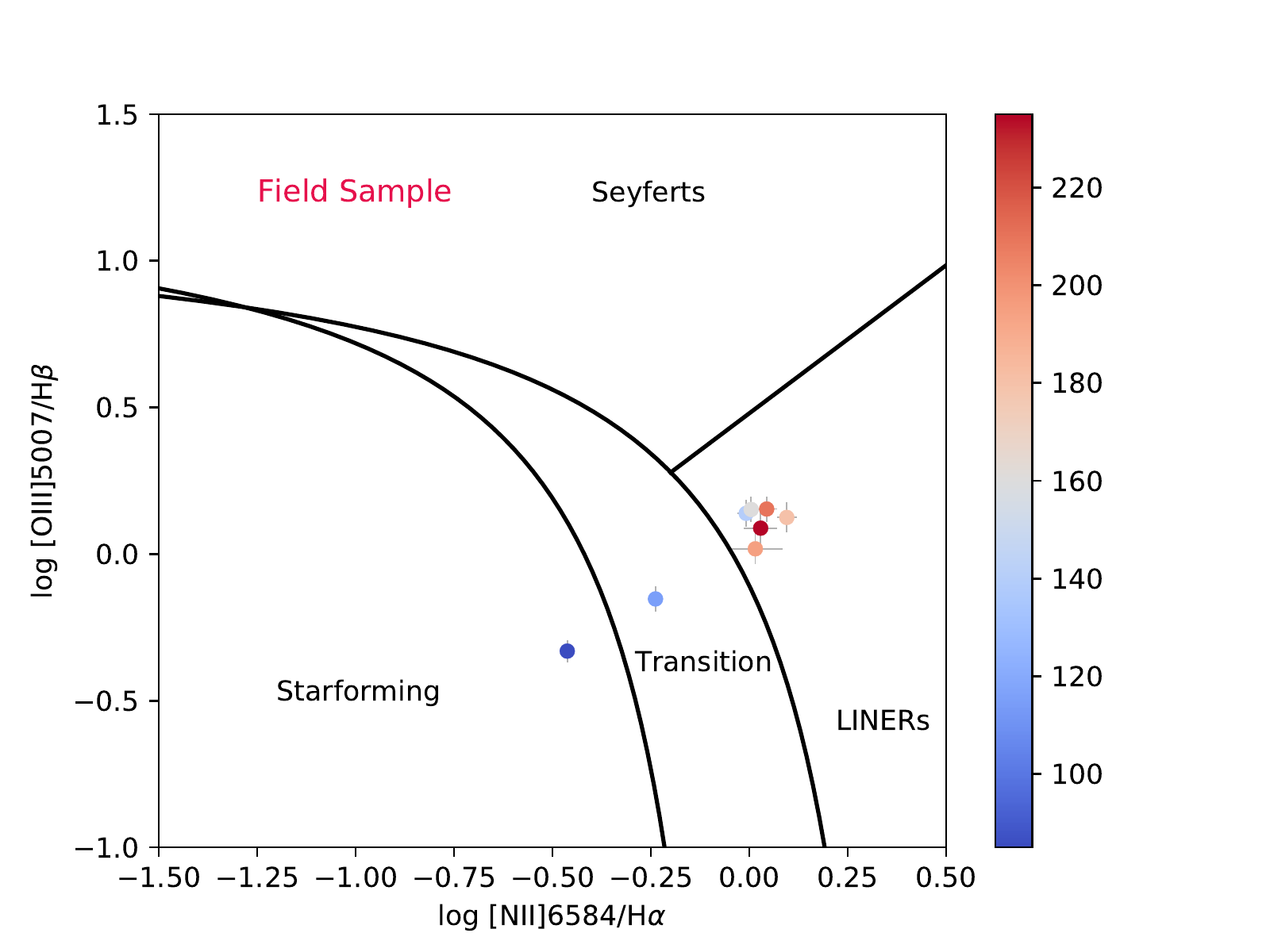}\\
 \end{tabular}
 \caption{The BPT diagram for the stacked spectra from the sample of ETGs in GCs (left) and in the field (right). The color scale is based on the velocity dispersion of the stack from blue to red as the velocity dispersion increases.}
 \label{fig:BPT}
\end{figure*}

For the BPT diagram \citep{BPT}, we use the AGN, Star Forming and Transition (Composite) separation lines defined in \cite{Kewley2001} and \cite{Kauffmann2003} and the limits set in \cite{Kewley2006} to separate Seyferts and LINERs. In Figure \ref{fig:BPT}, we show the BPT diagrams for the stacked spectra from the CGs and the field sample. The colours of the points indicates the galaxy velocity dispersion ($\sigma$), from dark blue (low-$\sigma$) to dark red (high-$\sigma$). The stacked spectra of ETGs in CGs is spread between the LINERs and Transition regions with the highest $\sigma$ stacks concentrated in the LINERs partition. For the field sample, two of the lowest-$\sigma$ stacks resides outside the ``LINERs'' part of the diagram where the other stacks are located. The stack with $70 \leq \sigma <100$ km/s is classified as ``Starforming'' and the stack with $100 \leq \sigma < 130$ km/s is classified as a ``Transition'' object.

The BPT diagram, albeit a good diagnostic regarding the ionization source
for galaxies with strong emission lines, is not able to discriminate between
genuine, AGN induced LINER-like emission and other ionization mechanisms unrelated
to accretion by a supermassive black hole. The WHAN diagram, on the other hand,
supplies a classification scheme for weak emission line galaxies whose classification is ambiguous, separating true LINERs from spectra whose emission lines are due to
ionization by hot, evolved low-mass stars (HOLMES) -- i.e. ``retired'' objects, with
no star formation whatsoever. The WHAN diagram therefore discriminates galaxies in five classes of gas ionization mechanisms using the following criteria:

\begin{enumerate}
\item pure star-forming galaxies: $\log[\mbox{NII]/H}\alpha < -0.4 $ and $W_{\mbox{H}\alpha} >$ 3 \AA\,
\item Seyferts: $\log[\mbox{N II]/H}\alpha > -0.4 $ and $W_{\mbox{H}\alpha} >$ 6 \AA\,
\item LINERs: $\log[\mbox{N II]/H}\alpha > -0.4 $ and $ 3 < W_{\mbox{H}\alpha} <$ 6 \AA\,
\item Retired galaxies: $W_{\mbox{H}\alpha} <$ 3 \AA\,
\item Passive galaxies: $W_{\mbox{H}\alpha}$ and $W_{[\mbox{N II}]} < 0.5$ \AA\, 
\end{enumerate}

\begin{figure*}
\centering
\begin{tabular}{cc}
\includegraphics[width=.52\textwidth]{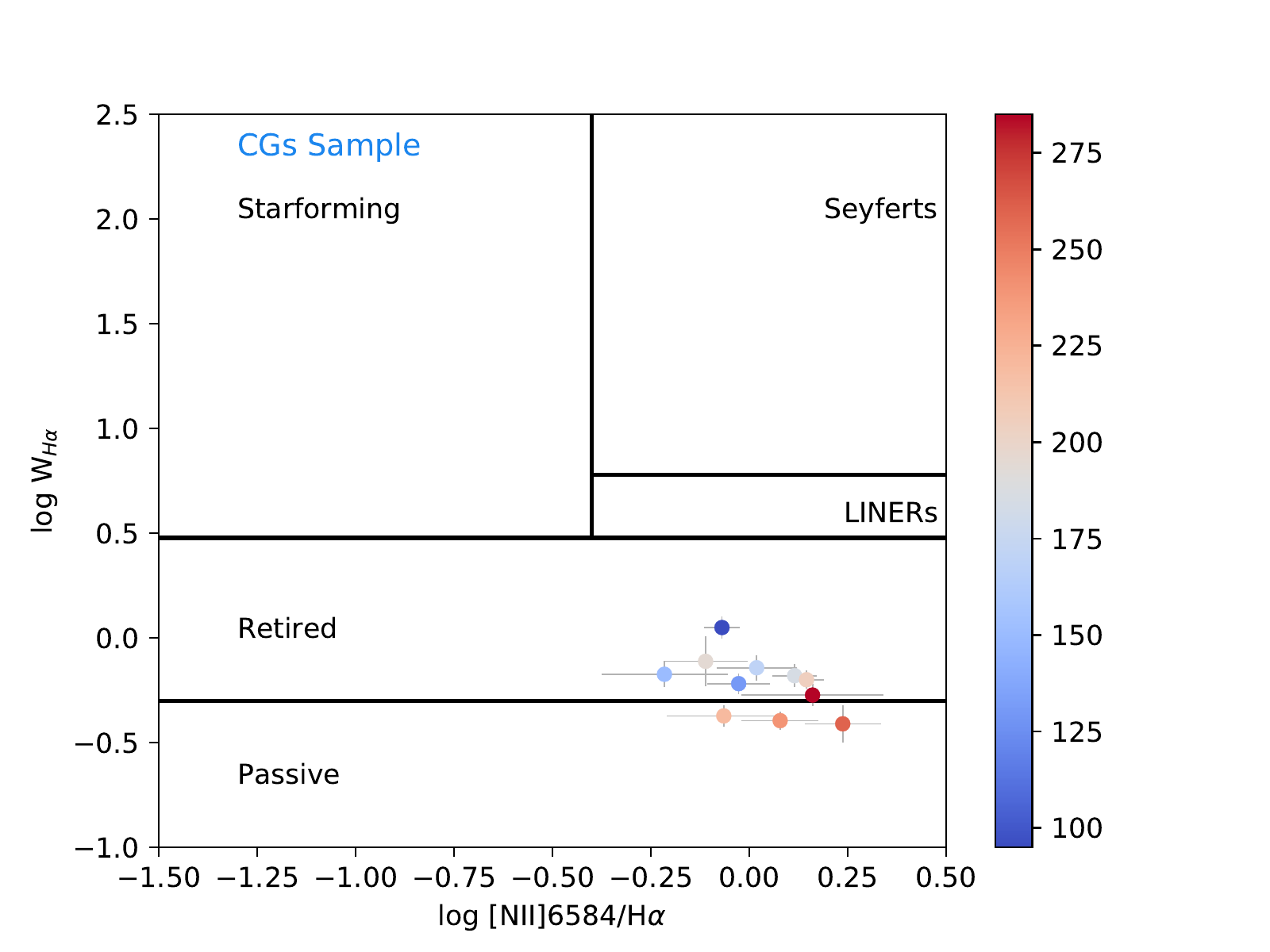} & \includegraphics[width=.52\textwidth]{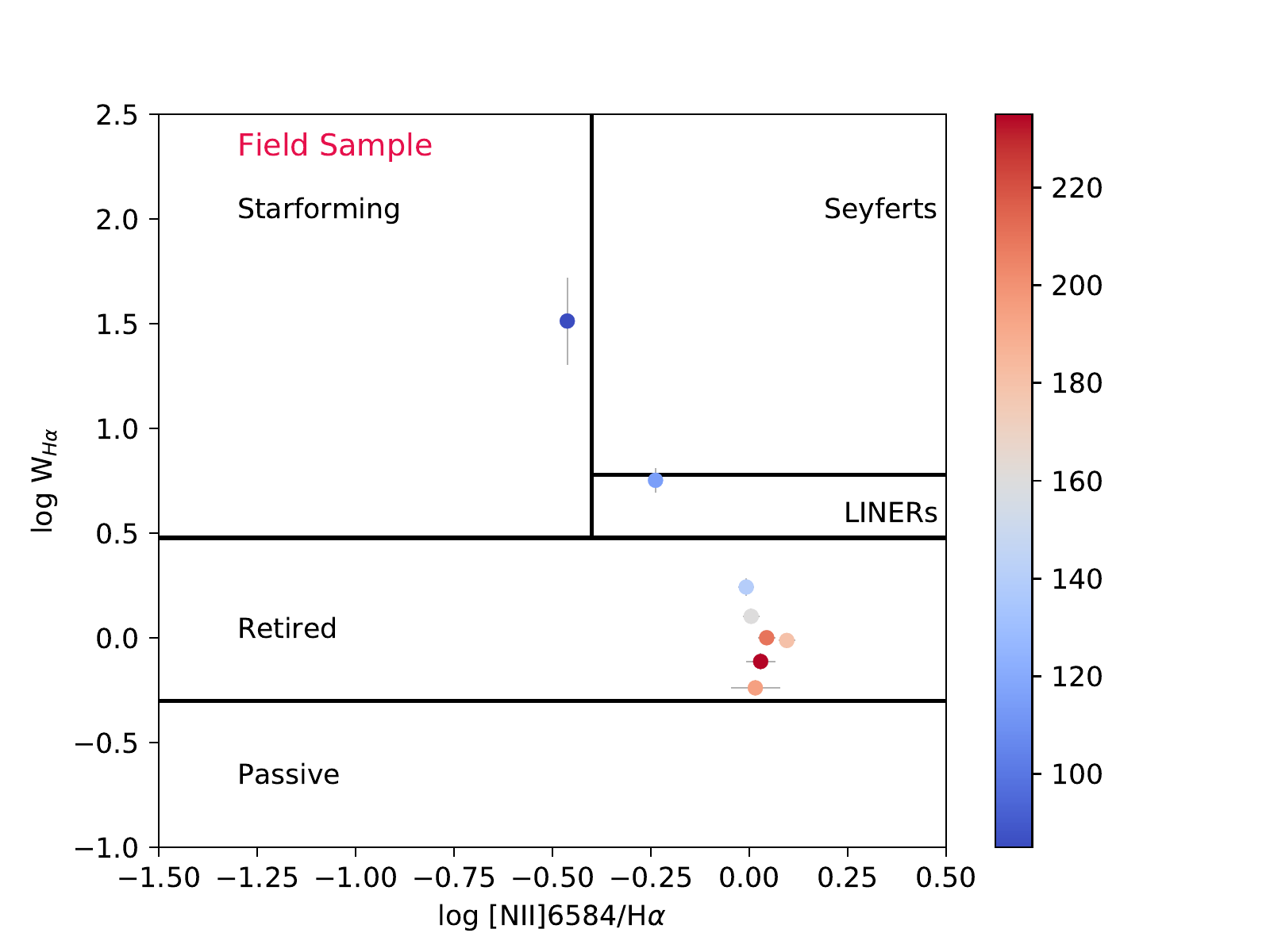}\\
 \end{tabular}
 \caption{WHAN diagram for the stacked spectra from the sample ETGs in GCs (left) and in the field (right). The color scale is identical to Figure~\ref{fig:BPT}.}
 \label{fig:WHAN}
\end{figure*}

In Figure \ref{fig:WHAN} we present the WHAN diagram for the stacked spectra from both samples. The stack spectra of ETGs in CGs with the highest $\sigma$ falls in the "Passive" region of the diagram while other stacks are mostly concentrated in the bottom part of the "Retired" area. The field sample is majority located in the ``Retired'' part of the diagram, with the lowest-$\sigma$ bins being closed to the LINERs area and the highest-$\sigma$ spectra are in the bottom of the ``Passive'' region. The exception is the two lowest $\sigma$ bins that is classified as Starfoming and LINERs. Notice, however, that the WHAN diagram presents a less defined frontier between Starforming and Seyfert-like spectra, so we can confidently confirm that the source of gas ionization is associated to young massive stars.

 Considering that our sample is composed of ETGs, the absence of active star formation as indicated by the BPT and WHAN diagrams is not a surprise. However, star formation \citep{patton+13} or AGN activity \citep{silverman+11} can be induced by galaxy interactions, resulting in an increase of AGN or SF optical signatures. The diagnostic diagrams have shown that the ionization source is similar for ETGs in CGs or in the field, with no detectable
contribution of ongoing star formation or ionization by a supermassive black hole in almost all bins. Therefore, the ionization field of the ETGs in our sample seems to present a small sensitivity to the environment. However, Figure \ref{fig:WHAN} also reveals a shift in the equivalent width of H$\alpha$ of field galaxies with relation to galaxies in CGs: even though the ionization agent does not vary significantly, the line emission is more intense in field galaxies. This hints at a reduction of the total ionizable gas budget in CG galaxies as compared to their field counterparts.

We have checked the robustness of our gas ionization source characterization scheme against stack contamination by galaxies containing strong emission lines. We have performed a visual analysis of individual spectra in a given stack in order to identify such objects. We have found a low ($\sim 9-13 \%$) level of contribution by strong line emitters. Excluding these objects from the stacks, the resulting equivalent widths of H$\alpha$ are reduced by 2-4\%. This variation is much lower than the typical differences between stacks and barely affects the position of a stack in the WHAN diagram. This very low sensitivity to outliers is due to our choice of median-combining rather than average-combining individual spectra. Our results are therefore not affected at all by objects with intense emission lines.

 \section{Dynamical Analysis}\label{sec:dynamic}

\begin{figure}
\centering
\includegraphics[width=0.50\textwidth]{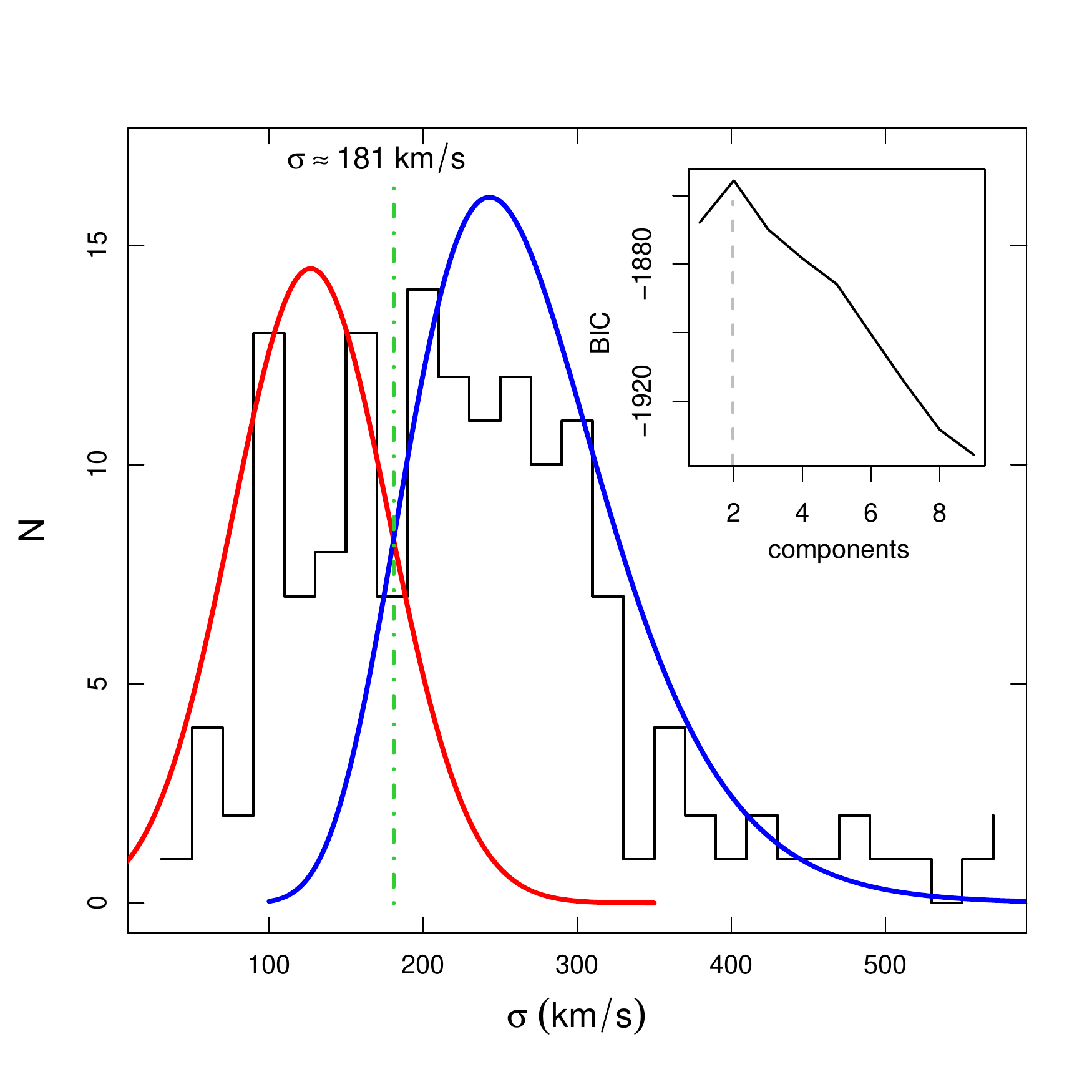}
\caption{The result from 1000 re-samplings and the fitting of a normal (red) and normal-lognormal (blue) in the mixture model. The CGs of our sample is divided into two groups concerning the velocity dispersion: low $\sigma$ groups with $\sigma_{G} \leqslant 181$ km/s and high $\sigma$ groups with $\sigma_{G} > 181$ km/s for $\sigma_{G}$ as the velocity dispersion of the group. We also show the Bayesian Information Criterion (BIC) that indicates that the distribution of velocity dispersion of the groups has two components.}
\label{fig:dyn_mclust}
\end{figure}

\begin{figure*}
\centering
\includegraphics[width=0.53\textwidth]{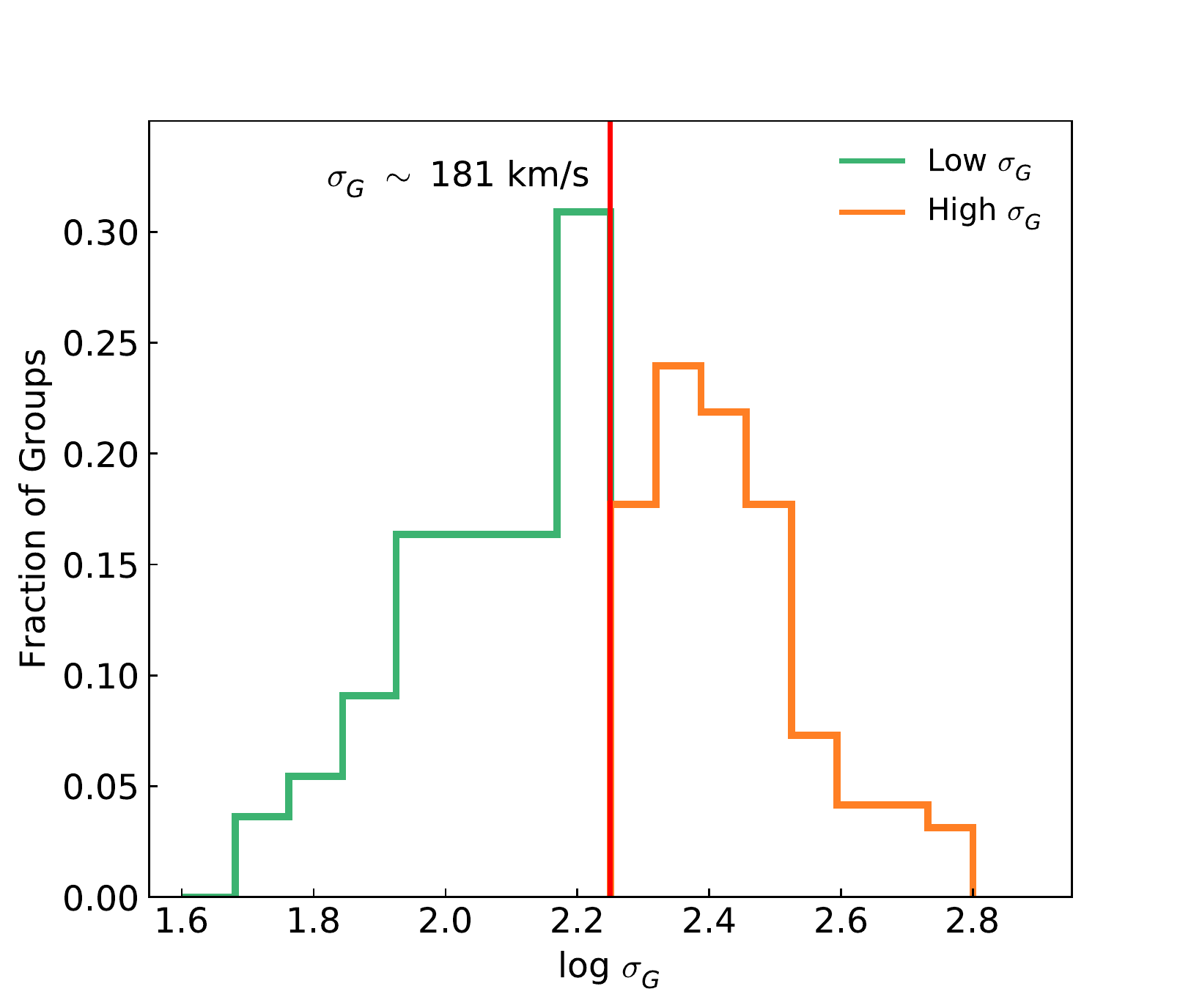}
\caption{Distribution of the 151 CGs of our sample in term of group velocity dispersion ($\sigma_{G}$). The groups are separated into two families : low-$\sigma$ and high-$\sigma$. The low-$\sigma$ groups have $\sigma_{G} <$ 181 km/s. The red dashed line is the threshold limit  that separates the two groups.}
\label{fig:din_logsig}
\end{figure*}

We investigate the dynamics of the CGs in our sample by performing an analysis of the velocity dispersion distribution using the MCLUST package for model-based clustering. MCLUST is an efficient  R package for modeling data as a Gaussian finite mixture \citep{FraleyAndRaftery2002}. A basic explanation of how MCLUST works is presented in \cite{Decetal2017}. In the present analysis, we initially ran the code for a Gaussian mixture model and found two modes in 97\% of the times out of 1000 re-samplings. This methodology shows the robustness of the finding. However, while MCLUST indicates bimodality in the data, the Gaussian mixture is not necessarily the best fit for the distribution. Taking this into account, we compare three specific mixtures: normal-normal, normal-lognormal, and normal-gamma distributions. We find the normal-lognormal mixture to be the best model from the likelihood ratio. Adopting this model, we divide our sample into two classes following the distribution of the velocity dispersion of the group ($\sigma_{G}$): low $\sigma$ groups  ($\sigma_{G} \leqslant 181$ km/s) and high $\sigma$ groups ($\sigma_{G} > 181$ km/s) as is shown in Figure \ref{fig:dyn_mclust}. The fraction of groups, from the total of 151 CGs, that falls in each regime is given in Figure \ref{fig:din_logsig}.

\begin{figure*}
\centering
\includegraphics[width=0.53\textwidth]{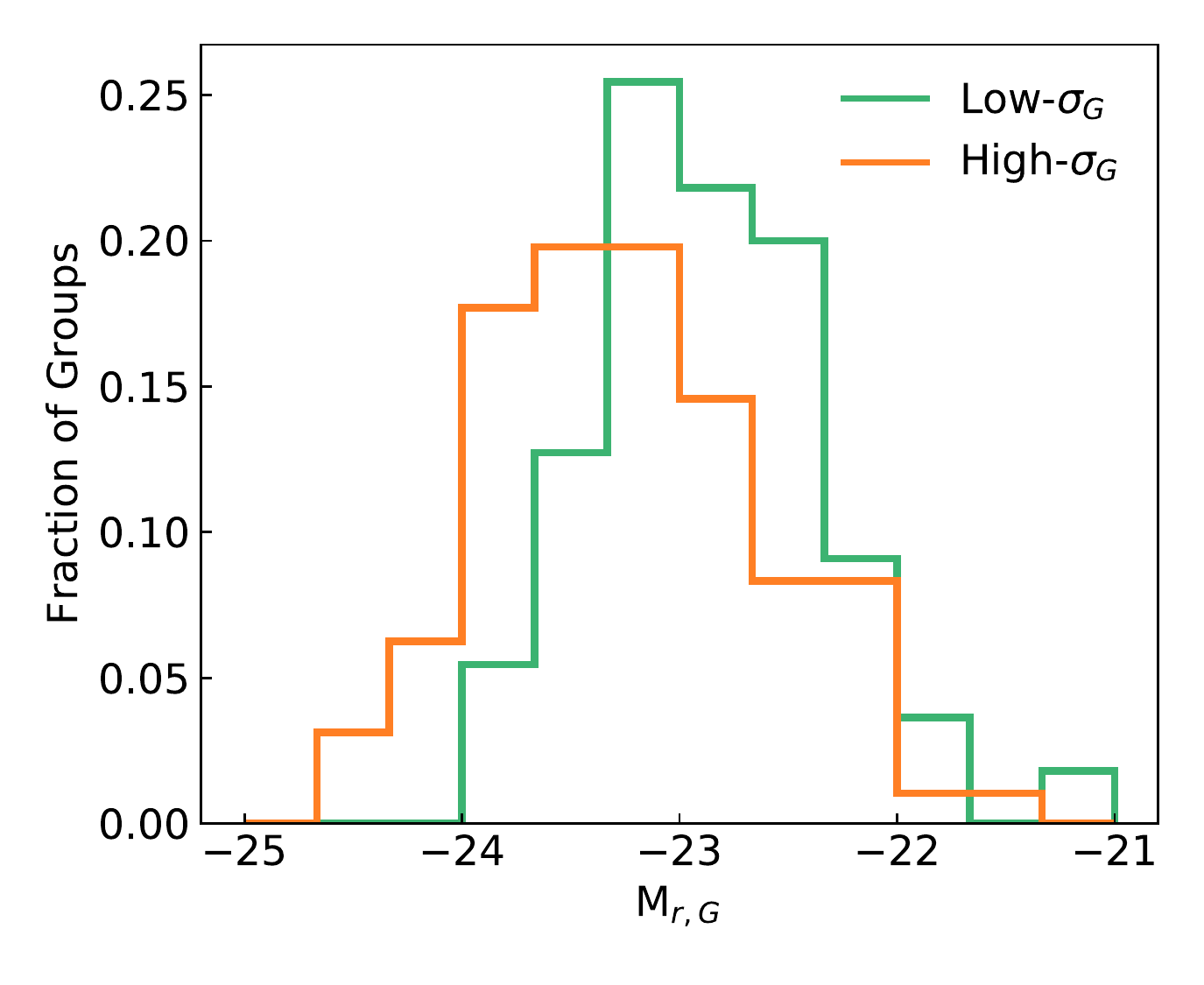}
\caption{The distribution in M$_{r}$ for the low and high $\sigma$ groups of our sample of CGs. The p-value calculated from the Permutation Test (p = 0.004) indicates that the distributions came from different parent populations.}
\label{fig:din_Mr}
\end{figure*}

The high and low $\sigma$ groups exhibit different absolute magnitude distributions (total luminosity of all galaxies in a given group) as we can see in Figure \ref{fig:din_Mr}. The permutation test\footnote{Using the function permTS in R package under the library perm \cite{Fay09}} takes all possible combinations of group membership and creates a permutation distribution from which one can assess evidence that the two samples come from two different populations. The test gives a p-value of 0.004 implying rejection of the null hypothesis that the two distributions come from the same parent population. Groups in the high-$\sigma$ regime have more luminous galaxies than the low-$\sigma$ groups with magnitudes extending to $M_{r}$ = -25 mag. 

\begin{figure*}
\centering
\includegraphics[width=0.53\textwidth]{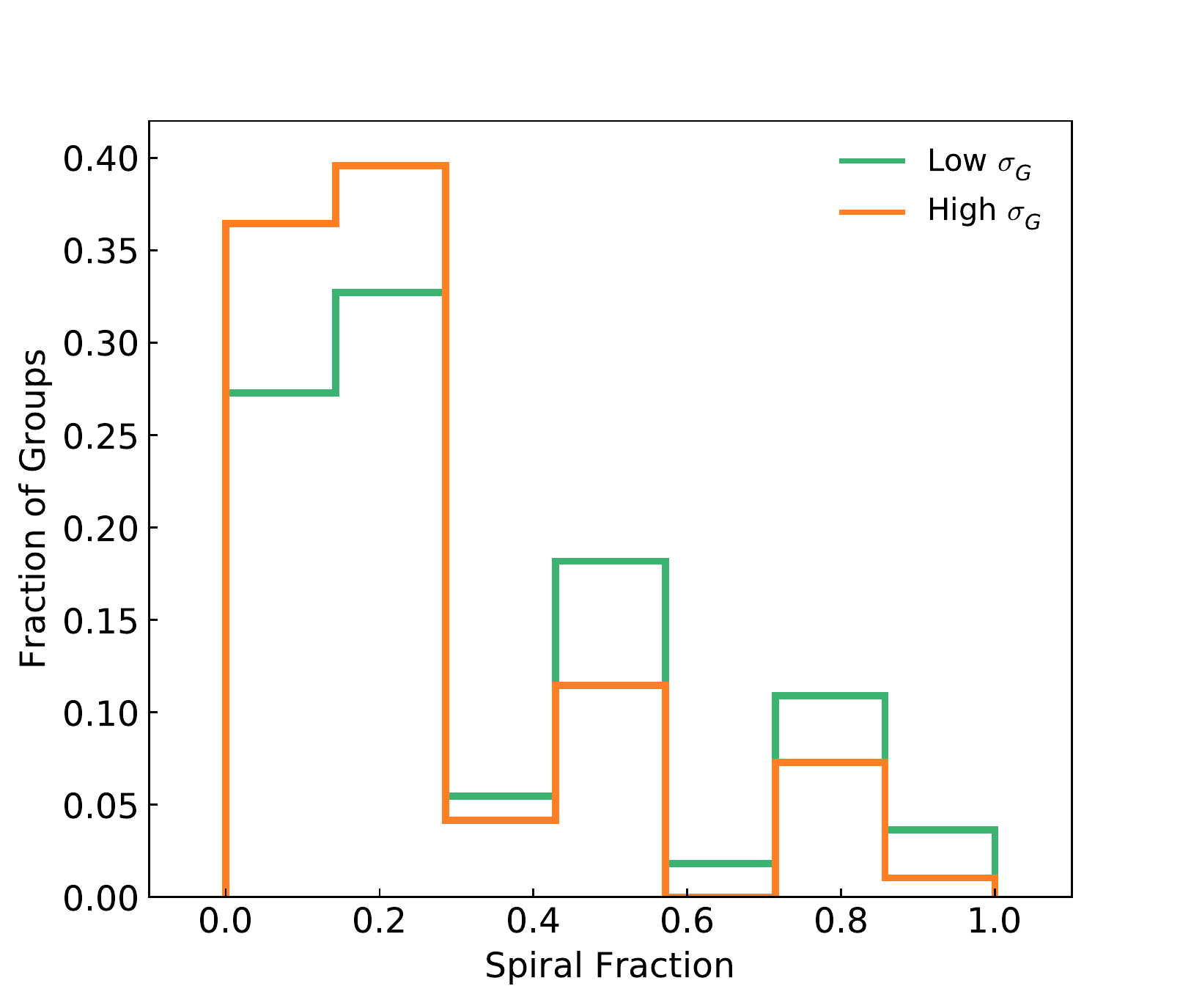}
\caption{The spiral fraction distribution for low and high $\sigma$ groups of our CGs sample. The p-value estimated from the Proportion Test (p = 0.03) confirms that the two distributions are distinct.}
\label{fig:din_sigma_morph}
\end{figure*}

\begin{figure*}
\centering
\includegraphics[width=0.55\textwidth]{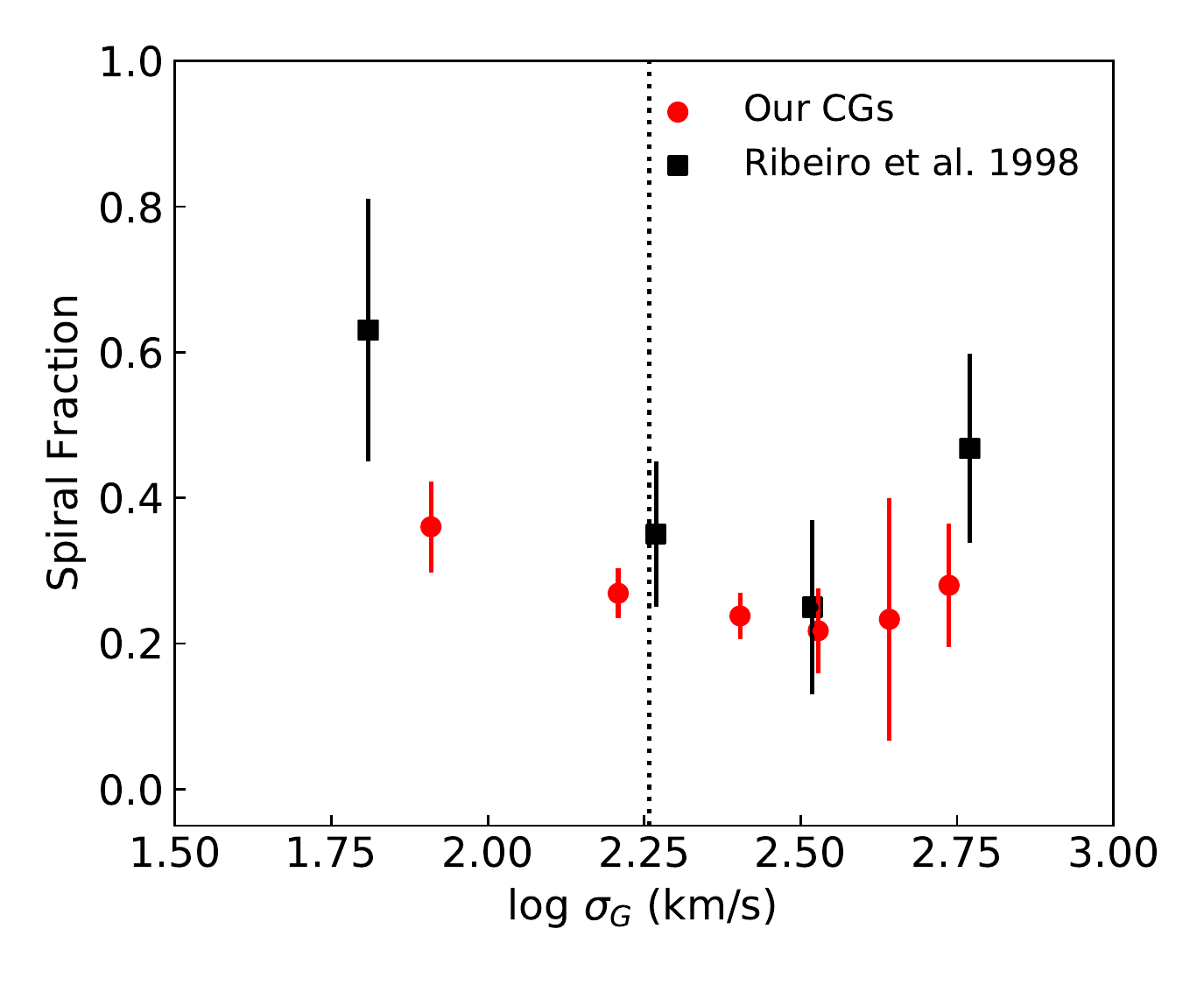}
\caption{The spiral fraction against the velocity dispersion for CGs of our sample (red dot). The traced line separates the two velocity dispersion groups. We also show the results from the study by \citet{Ribeiro98} with 17 CGs (black square).}
\label{fig:din_logsigxfracsp}
\end{figure*}

The two groups also distinguish themselves {\it wrt} the predominant morphological types. In Figure \ref{fig:din_sigma_morph}  we show the distribution of the spiral fraction for low and high-$\sigma$ groups where
we can clearly see how different the distributions are, confirmed by the p-value estimated using the Proportional Test\footnote {Using the function {\it prop.test} in R package} (p-value of 0.03), for testing how probable it is that both proportions are the same. The fraction of groups with low spiral fraction ($<$0.3), namely dominated by early-type systems, is larger for high-$\sigma$ groups - more massive groups tend to have more early-type systems. On the other hand, examining the fraction of groups with high spiral fraction ($>$0.3), we conclude that less massive groups are dominated by late-type galaxies. In Figure \ref{fig:din_logsigxfracsp}, we exhibit the fraction of spirals versus group velocity dispersion relation where again the clear correlation (spiral fraction decreases with group velocity dispersion) is noticeable. We note that the last bin (higher velocity dispersion) shows an increase in spiral fraction, result also obtained by \citet{Ribeiro98}. We also show in Figure \ref{fig:din_logsigxfracsp} the result from \citet{Ribeiro98} in the study of dynamical properties of 17 HCGs. In \citet{Ribeiro98} the morphological classification is based on the equivalent width of the H$\alpha$ spectral line ($\rm EW(H\alpha) > 6 \AA\,$ is considered a late-type) which explains the slightly higher spiral fraction when compared to ours. Nevertheless, both trends are quite in agreement.

\begin{figure*}
\centering
\includegraphics[width=0.55\textwidth]{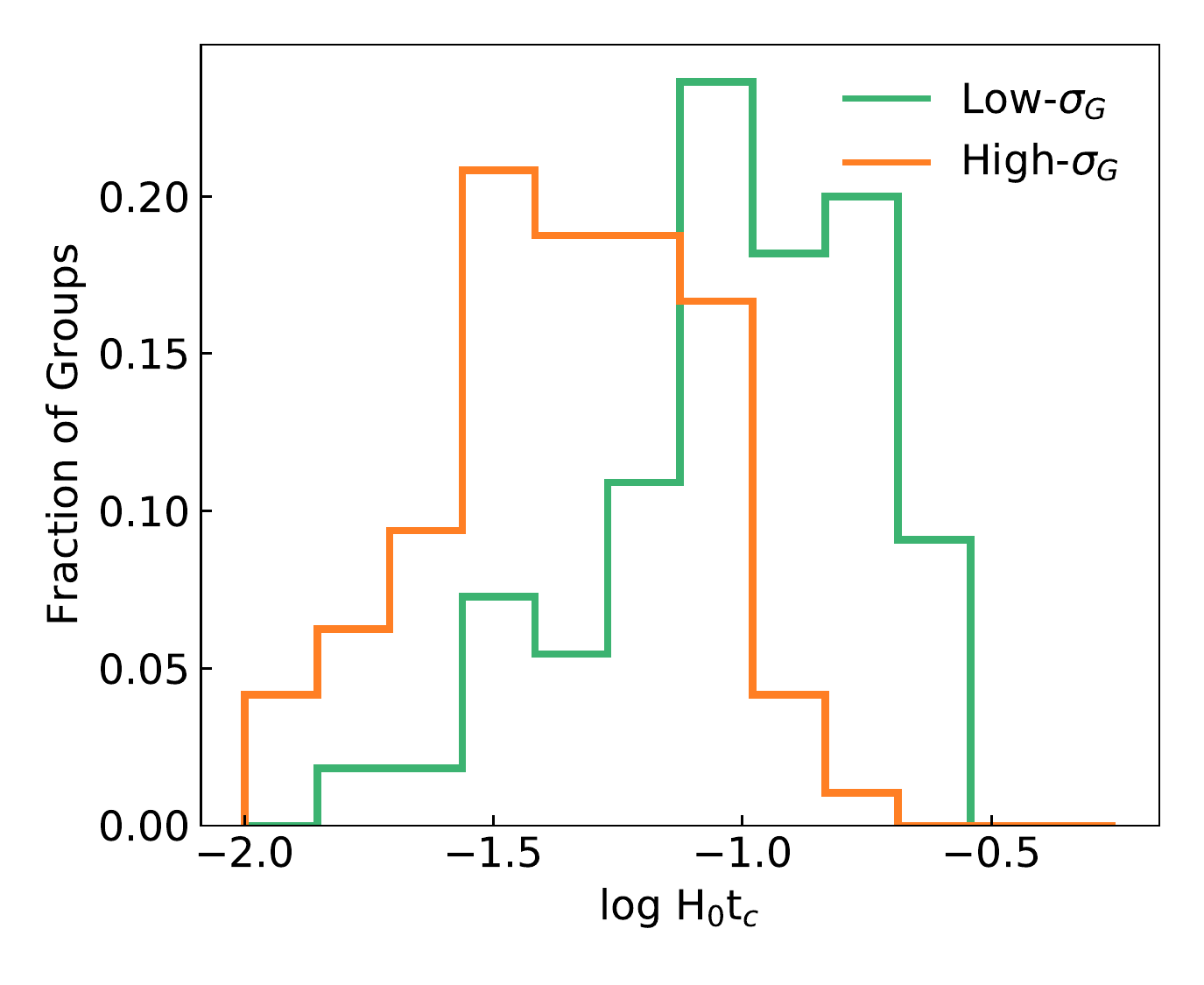}
\caption{The distribution of the crossing time for low and high-$\sigma$ groups of our sample.}
\label{fig:din_sigma_tc}
\end{figure*}

An additional parameter revealing the CGs dynamics is the crossing time, defined as the time for a galaxy transverse the group. Its simple version can be written like:

\begin{equation}
t_{c} = \frac{4}{\pi}\frac{R}{V}
\end{equation}

\noindent where $R$ is median of the two-dimensional galaxy-galaxy separation vector and $V$ is the three-dimensional velocity dispersion estimated as $V = [3(\langle v^2\rangle - \langle v \rangle^2 - \langle \delta v^{2}\rangle)]^{1/2}$, with $v$ being the radial velocity of the galaxies of the group, $\delta v$ the velocity error and the bracket means the average over all galaxies in the group. From the distribution of the crossing time for low- and high-$\sigma$ groups, shown in Figure \ref{fig:din_sigma_tc}, we can clearly see that low $\sigma$ groups have higher crossing times.

\begin{figure*}
\centering
\includegraphics[width=0.55\textwidth]{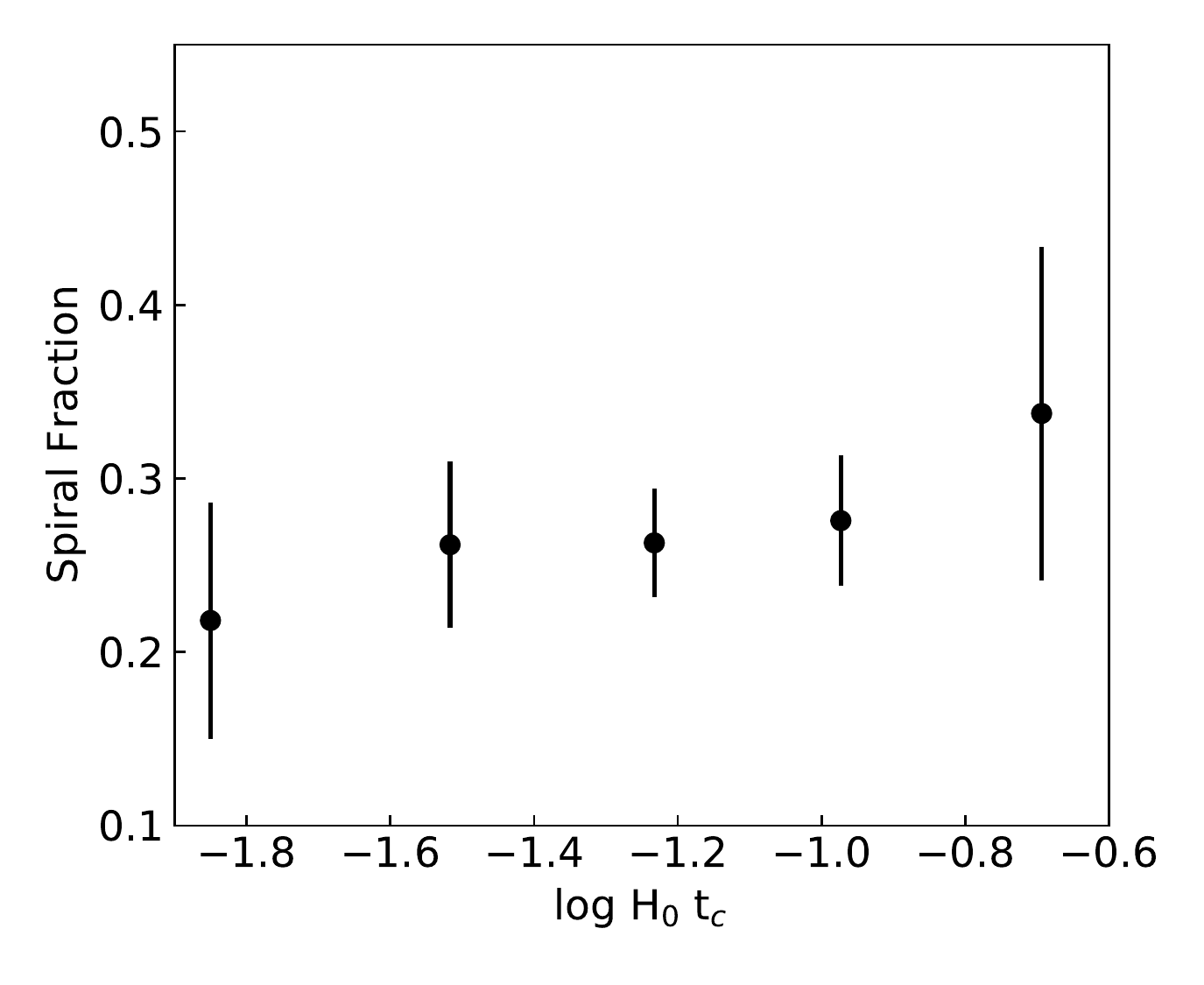}
\caption{The fraction of spiral as a function of the crossing time for the 151 CGs of our sample. The points are given by the mean in each bin and the error bar is the 1-$\sigma$ error.}
\label{fig:din_logtcxfracsp}
\end{figure*}

It is expected that groups with a small crossing time will suffer more galaxy-galaxy interactions. Since such interactions are responsible for transforming the morphology of a galaxy, from late-type to early type, it is reasonable to expect that groups with smaller crossing times show a lower fraction of late-type galaxies. In Figure \ref{fig:din_logtcxfracsp} we show the fraction of spirals as a function of the crossing time for 151 CGs in our sample. The points represent the mean of the crossing time of at least 11 groups in each bin and the error bar is the 1-$\sigma$ error. We found the same correlation as \cite{Hickson92} and \cite{Ribeiro98}: the spiral fraction is is lower for groups with small crossing time.

\begin{figure*}
\centering
\includegraphics[width=1.0\textwidth]{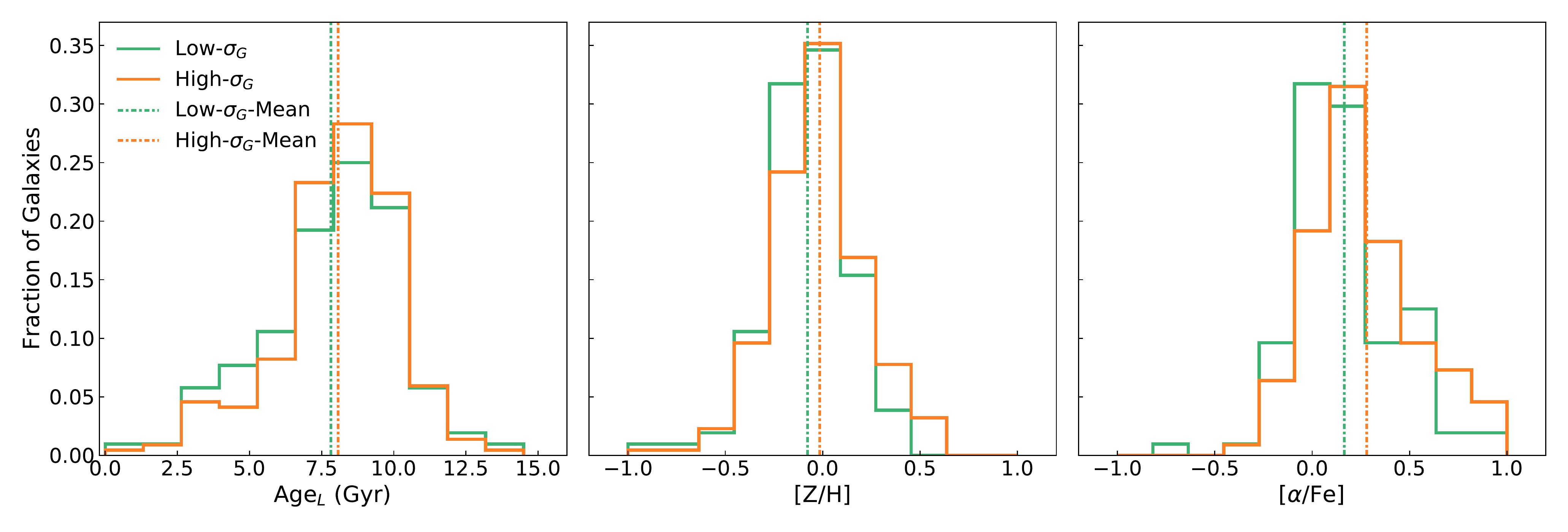} 
    \caption{The stellar population parameters of the ETGs in the high and low-$\sigma$ groups. The dashed line is the mean values of the parameter. The p-value calculated for the distribution of the parameters [Z/H] (p = 0.027) and [$\alpha$/Fe](p = 0.001) in the Permutation Test suggests the rejection of the null hypothesis which claims that the parameters distributions are from the same parent population. For the age parameter, the p-value = 0.341 it is an indication that the distributions are similar. \label{fig:dyn_stellar}}
\end{figure*}

Another important aspect of the study of CGs is to understand how their dynamical properties are linked to the stellar population properties of the member galaxies. In Table \ref{tab:dynamics} we list the dynamical parameters for all the 151 CGs in our sample. The parameters are: (1) the identification of each group; (2) number of members in each group; (3) total absolute magnitude of the group in the r-band, M$_{r,G}$; (4) velocity dispersion of the group; (5) harmonic radius (Mpc);  (6) total dynamical mass (in $M_{\odot}$); (7) the crossing time (in H$_{0}^{-1}$); (8) spatial  density; (9) spiral fraction; and (10) dynamical class, either low-$\sigma$ (L) or high-$\sigma$ (H). In Figure \ref{fig:dyn_stellar}, we present the distributions of Age, [Z/H], and [$\alpha$/Fe] for ETGS belonging to low and high-$\sigma$ groups. Additionally, we estimate the mean value of each distribution for the parameters: Age(L, H)$=(7.82, 8.07)$ Gyr; [Z/H](L, H)$=(-0.07, -0.01)$; and [$\alpha$/Fe](L ,H)$=(0.16, 0.28)$.  Although the mean values are close, between low and high-$\sigma$ distributions, the Permutation Test indicates that the [Z/H] and [$\alpha$/Fe]
parameters have different distributions in low- and high-$\sigma$ CGs, with p-values equal to 0.027 and 0.001, respectively. On the other hand, the age distribution for high and low-$\sigma$ groups are almost indistinguishable as is clear from the median values and the p-value = 0.341.

Looking at the distributions and mean values of the age, [Z/Fe], and [$\alpha$/Fe] of Figure \ref{fig:dyn_stellar}, it can be noted that the ETGs belonging to the high$-\sigma$ CGs are formed slightly earlier and faster than the low$-\sigma$ CGs. The age difference in the mean values between the high-$\sigma$ and low$-\sigma$ is almost negligible around 0.25 Gyr, and from the histograms in age, we can see that there are more low$-\sigma$ CG members with age between $\sim2.5$ and 6 Gyr ($\sim 20\%$ of the galaxies). For the mean value of [$\alpha$/Fe], the high-$\sigma$ members are $\sim 0.12$ dex more enhanced than the low-$\sigma$ ones. This is in agreement with the low-$\sigma$ CGs were formed recently compare with the high$-\sigma$ CGs. A more detailed dynamical study for a better understanding of the formation of CGs is required.

\section{Discussion and Summary}\label{sec:discussion}
 
\subsection{Stellar Population Parameters}

We investigate the stellar population present in ETGs in the high and low-density environments given by the CGs and field, respectively. Significant differences in the age, metallicity and $\alpha$-enrichment of those populations between these two regimes are expected, since the CGs environment is a very favourable environment for interactions, such as mergers. Previously,  \cite{deLaRosa07} suggested that there is a truncation in star formation of ETGs in CGs based on the behaviour of  [$\alpha$/Fe] versus central velocity dispersion ([$\alpha$/Fe] is smaller for larger velocity dispersion -- Figure 8). It is important to keep in mind that these results come from a small sample of 22 ETGs in Hickson Compact Groups. However, as presented in Section \ref{subsec:hybridmet} and in Figure \ref{fig:hybrid_CGxField}, our results do not support such interpretation, since  $[\alpha/Fe]$ was shown to increase for higher central velocity dispersion.

Motivated by the morphology-density relation \citep{Dressler80} and the Butcher-Oemler effect \citep{BO78}, we compare the properties of ETGs in CGs with those in other environments probing a large domain in spatial density. \cite{SPIDER10} studied a sample of 20,977 bonafide central and satellites ETGs as given in the catalog of \cite{Yang07}. From the definition adopted, the central are those with the highest stellar mass in the group. The central sample was divided based on the host halo mass (lower and higher than log($M_{h}$/$M_{\odot}) = 12.5$ ) while the satellite sample was divided into three parts: 1) those inhabiting a low mass halo ($\log(M_{h}/M_{\odot} < 14$); 2) those in massive haloes ($\log M_{h}/M_{\odot} \geq 14$); and 3) satellites in the outskirts of groups ($R >$ 0.5 $R_{200}$). The analysis of the stellar population parameters for both samples indicates that only the central galaxies have a dependency with the environment, where central located in high mass halos display younger ages than central in lower mass halos. The satellites show no correlation with the environment except for galaxies in the low-velocity dispersion regime.

\begin{figure*}
\centering
\includegraphics[width=0.48\textwidth]{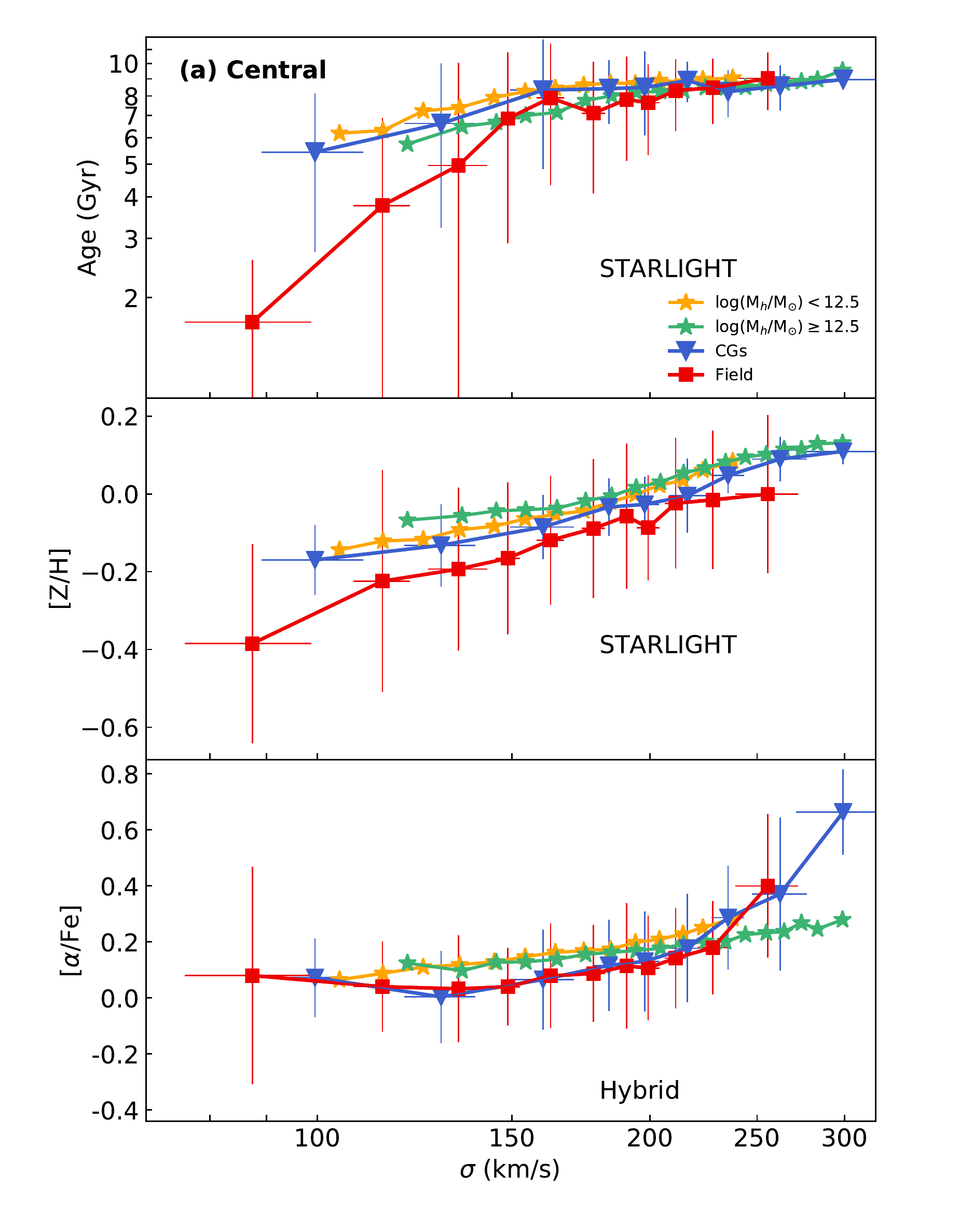} 
\includegraphics[width=0.48\textwidth]{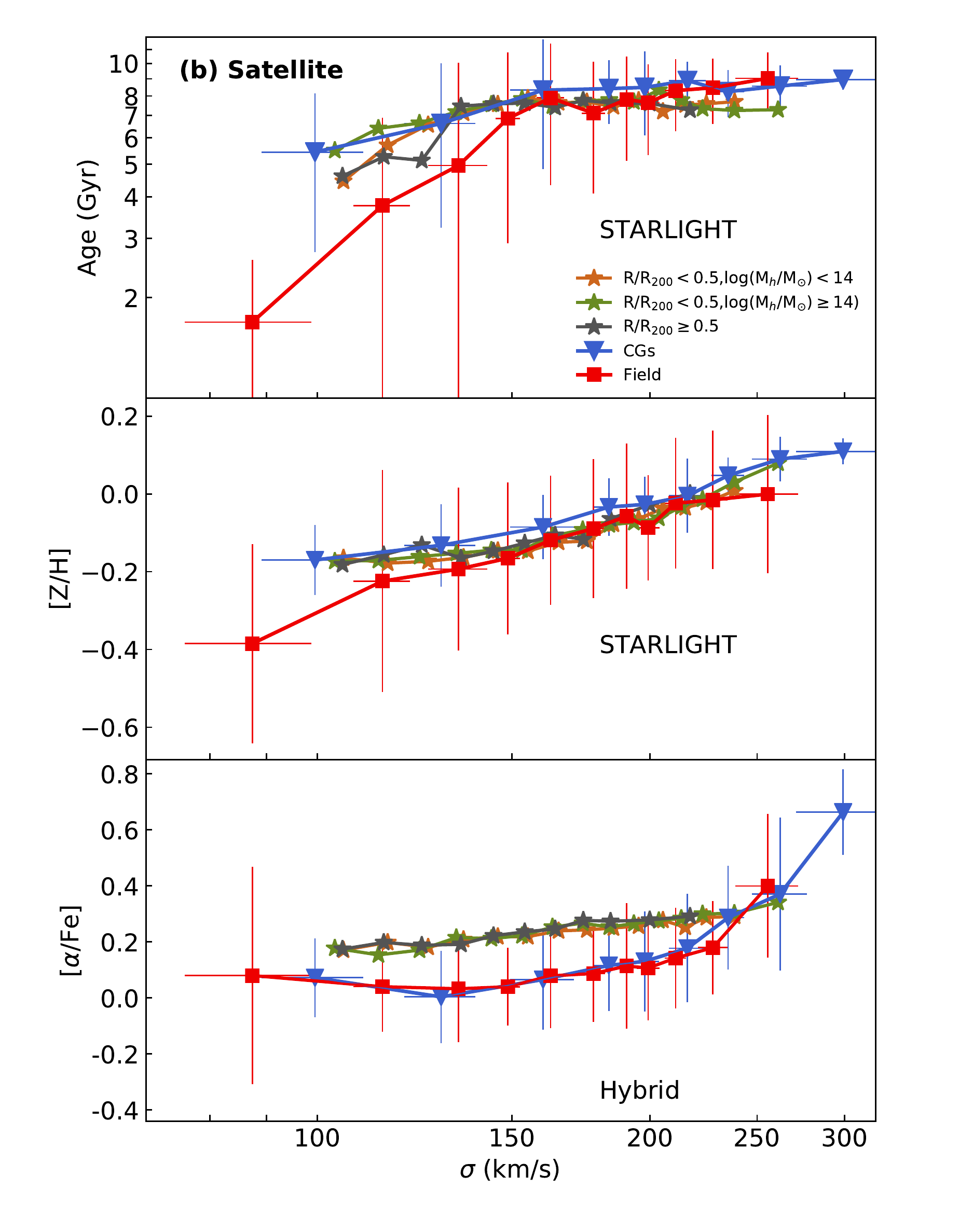}
\caption{Comparison between the stellar population parameters for central (a) and satellites ETGs (b) defined in \citet{SPIDER10} and our sample of ETGs in GCs and
the field. In \citet{SPIDER10} the environment is defined by the dark matter halo mass and the central are those galaxies with
higher stellar masses from the \citet{Yang2007} catalog. In panel (a) the central embedded in low halo mass ($\log M_{h}/M_{\odot} < 12.5$) are consider the ``isolated'' central while the central in massive halo ($\log (M_{h}/M_{\odot} \geqslant 12.5$) are the ``group'' central. For panel (b), the satellite sample are divided in three sub-samples: satellites embedded in low mass halo ($R/R_{200} < 0.5 \log M_{h}/M_{\odot} < 14$), massive haloes ($R/R_{200} < 0.5 \log M_{h}/M_{\odot}\geqslant 14$) and satellites in more external regions of the group ($R/R_{200} \geqslant 0.5$). To compare with the results from \citet{SPIDER10}, we use the age and [Z/H] from the spectral fitting and the [$\alpha$/Fe] from the hybrid method.}
\label{fig:comp_spiderx}
\end{figure*}

In Figure \ref{fig:comp_spiderx} (a), we contrast the stellar population properties of ETGs in CGs with those of field, satellites and centrals in low and high halo mass systems. For a consistent comparison with the results from \citet{SPIDER10}, we estimate the luminosity-weighted metallicity using the spectral fitting approach (Equation \ref{eq1}). The age and [$\alpha$/Fe] parameters are estimated for all the environments following the hybrid method mentioned in Sec. 3.4. In general, all the stellar population parameters increase with the velocity dispersion, confirming previous findings that velocity dispersion is the main driver of the stellar populations properties. We also see this trend as a manifestation of the downsizing scenario where massive galaxies formed their stellar content in remote epochs and currently star formation is happening in the low mass systems. This is particularly true for $\sigma \leq 150$ km/s, where it is clear that field ETGs are considerably younger than centrals and ETGs in CGs ($\Delta$ Age $\sim$ 2.3 Gyr). This behaviour is in agreement with \cite{Thomas10} who show that the low mass ETGs are more affected by the low-density environments. Age gets lower for centrals and satellites as well, but not as much as for Field ETGs. Regarding the [Z/H] parameter, there is a clear environmental effect with centrals in more massive halos being more metal-rich than field ETGs with the difference reaching $\Delta [Z/H] \sim 0.16$ dex. ETGs in CGs seem to fall in between these two classes. Finally, the behavior of [$\alpha$/Fe] parameter shows some interesting features. It is slightly higher for centrals when compared to Field and CG galaxies ($\sim$ 0.1 dex) up to $\sigma \sim 210$ km/s . However, for $\sigma \gtrsim 250$ km/s [$\alpha$/Fe] for ETGs in CGs increases by almost 0.4 dex. This very clear trend may be interpreted as a sign of truncation of the star formation and this may be due to dry merge happening at an early phase of CG formation. It is important to stress that this is the first time that a clear difference between properties of ETGs in CGs and other environments is found.

As far as the comparison to the satellite systems is concerned we are restricted to $\sigma$ between 100 - 250 km/s since for the extremes there is no corresponding data points in the satellite sample. In this central velocity dispersion range, the age behaves similarly as in the comparison with centrals, with ETGs in the Field being younger for $\sigma < 150$ km/s ($\Delta \rm Age \sim 4$ Gyr) . Satellite galaxies in the outskirts of groups (for $R/R_{200} > 0.5$) assume ages closest to the field sample as is expected since this sample represents an environment closer and closer to the Field galaxies (low density regime). As for the [Z/H] parameter, we find an increasing tendency for all environments, namely [Z/H] increases monotonically towards more massive galaxies. Notice, however, that Field ETGs tend to be more metal poor than ETGs in other environments. In the case [$\alpha$/Fe], we see a well established linear relation for satellites in all different groups, as previously found by \citet{SPIDER10}. For the velocity dispersion range of 100 to 250 km/s there is an offset of $\Delta [Z/H] \sim 0.1$ dex, which may be indicating the well-known quenching process acting in satellite galaxies whose consequence is the suppression of the star formation. This quenching mechanism may be caused ram pressure or frequent high-speed interactions with members of the group. 

Finally, it is important to emphasize that our samples of ETGs in CGs and in the Field, exhibit the same trend for the three stellar population parameters, with the exception for the age at the low-$\sigma$ regime and [$\alpha$/Fe] at the high mass end. The low-mass ETGs are younger in the low-density environment (Field) while more massive ETGs in CGs seem to suffer star formation truncation that leads to an increase in [$\alpha$/Fe] when compared to Field galaxies.

\subsection{Gas ionization}

Given the nature of CGs (moderate velocity dispersion and high number densities)  it is expected that galaxy interactions occur frequently and those interactions could trigger the activity in galactic nuclei by channeling the gas to the central region of the galaxies and enhancing star formation.  But it is still an open debate the influence of the interactions as a feeding process of SMBHs, and many works have indicated that the fraction of AGN in high density environments could even be smaller than in the less dense environments (ex:\cite{Dressler99} and \cite{Martini2007}).
Studying galaxies in HCGs, \cite{Coziol1998b}, \cite{Gallagher2008} and \cite{Martinez2010} estimated a higher fraction of AGN (between $41\%$ and $54\%$) in those groups compared to other environments. However, \cite{Sabater2012}, found no difference between the AGN fraction estimated for their sample of isolated galaxies and that estimated by \cite{Martinez2010} for galaxies in CGs. 

Based on the WHAN diagram, we find that the emission lines present in ETGs of our sample are due to the presence of HOLMES. This may indicate that dry merge in CGs is the main mechanism if merger is as effective as it is expected. Also, we see a clear trend in the sense that higher velocity dispersion ETGs are closer to the passive region in the WHAN diagram, implying a softening of the ionization field or a lower overall gas abundance.  However, we do not see
any significant contribution of ongoing star formation or ionization from AGN in CGs. On the other hand, an overall increase in H$\alpha$ emission is detected for the field sample, in particular at very low velocity dispersions. This implies that differences between low and high density regimes are not significant as far as the ionization field is concerned, but suggest a higher gas abundance in field galaxies in comparison to galaxies in CGs. One possible explanation for this feature is the occurrence of modest gas loss in CGs due to tidal interactions, which could preserve intact the stellar population properties. This would explain why we do not detect pronounced differences in mean stellar age, metallicity and alpha enrichment between galaxies in CGs and in the field.

\subsection{Dynamics}

\begin{figure*}
\centering
\includegraphics[width=.64\textwidth]{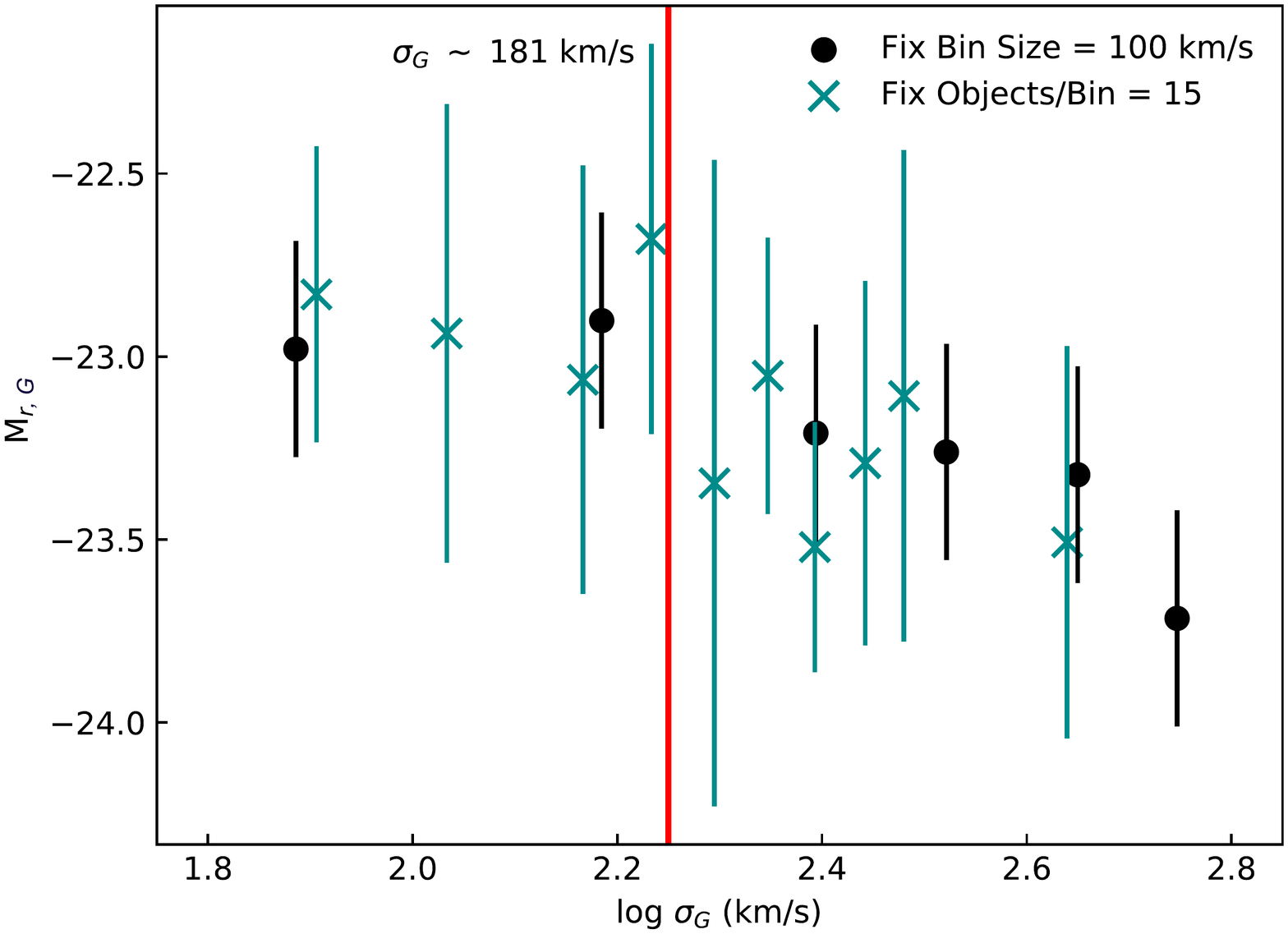}
\caption{The relation between the absolute magnitude of the group and the velocity dispersion. We use a fix size of bin ($\Delta$ = 100 km/s) and fix number of objects per bin (N/bin = 15) and in both cases the relation is linear.}
\label{fig:dyn_mrxlogsig}
\end{figure*}

The CGs have their dynamics associated with galaxy-galaxy interactions, mainly  merger process and tidal interactions. It is expected by the fast merger model \citep[]{Oliveira1994, Flechoso2001} that CGs become a giant elliptical galaxy as result of multiple merging events. In this sense, it is reasonable to assume that CGs contribute to the ETG field population. However, it is still a matter of dispute what is the timescale for the evolution of a typical CG. Our dynamical analysis indicate that our sample of 151 CGs may be separated into two groups according to their velocity dispersions (high and low $\sigma_{G}$ groups). We distinguish these two dynamical stages as follows: 1) the high-$\sigma$ groups as bound systems in virial equilibrium; and 2) the low- $\sigma$ groups as associations that probably formed more recently. The low-$\sigma$ groups could be taken as chance alignments \citep{Mamon2000}, but this not seem to be the case. The relation of the absolute magnitude of the groups versus the velocity dispersion shown in Figure \ref{fig:dyn_mrxlogsig} is linear for the whole $\sigma_{G}$ regime as expected for bound systems.

The high-$\sigma$ groups are more luminous and have more ETGs ($\sim 76\%$) than the low-$\sigma$ groups ($\sim 67\%$). We also find, as already stated in the works of \cite{Hickson88}, \cite{Ribeiro98} and \cite{Coziol2004}, that the fraction of spirals decreases for groups with higher velocity dispersion and that groups in the high-$\sigma$ regime have also smaller crossing times. Since these groups also have a low spiral fraction, it is reasonable to conclude that high-$\sigma$ groups are dynamically old structures. Another important piece of information reinforcing the presence of two families of CGs is the fact that the ETGs in these two families have exhibited stellar population properties significantly different.

\subsection{Summary}

We have performed a study of the galaxy stellar population parameters, the gas ionization and the dynamical properties of 151 CGs from the catalog defined by \cite{McCon09}. Our results can be summarized as follows:

\begin{enumerate}

\item The stellar populations in ETGs belonging to CGs present the same behavior as ETGs in the field following the analysis of the parameters Age, [Z/H] and [$\alpha$/Fe], indicating that spatial density is not responsible for establishing the stellar content of these systems. We do not confirm the truncation in star formation observed by  \cite{deLaRosa07}. It is worthy to note that the results shown by \citet{deLaRosa07} are based in a rather small sample composed of 22 ETGs in CGs. In our results, we find that  [$\alpha$/Fe] increases with central velocity dispersion of the ETGs in both environments, CGs and field.

\item Comparing ETGs in CGs with similar systems in low and high mass halos, we find essentially the same trend
of Age, [Z/H] and [$\alpha$/Fe] with central velocity dispersion. Considering that we are probing four orders of
magnitude in environmental density (from field to centrals), the similarity of these trends may imply a high regularity in the physical process that establishes the stellar population. 

\item Examining the behavior of Age as a function of central velocity dispersion we notice that in the low-$\sigma$
regime the less dense is the environment the younger is the stellar population in it. This suggests that quenching may depend on the environment -- systems with higher halo mass stop star formation more efficiently. On the other hand, looking at the high-$\sigma$ end ($>$ 250 km/s) we find that ETGs in CGs have [$\alpha$/Fe] $\sim$0.2 dex
greater than centrals inhabiting high mass halos, indicating truncation in star formation for the ETGs in CGs.

\item We identify in our sample of 151 CGs two main families: the low-$\sigma_{G}$ groups ($\sigma_{G} \leq 181$ km/s) with larger crossing times and higher fractions of spirals which can be interpreted as recently formed groups while the high-$\sigma$ groups, with smaller crossing times and smaller spiral fraction are those supposedly in virial equilibrium;

\item Analysis of the stacked spectra of ETGs in CGs and in the field have shown that these galaxies are located
in the same part of the WHAN diagnostic diagram. We see that the ionization source is similar for ETGs in CGs and in the field, with no detectable contribution of ongoing star formation or ionization by a SMBH for galaxies in CGs. Therefore, there are no differences between dense and low-density environments regarding the gas ionizating agent, but the overall gas emission is more intense in field galaxies than in their CG counterparts, hinting at a gas-loss mechanism operating in the CG environment, like tidal interactions.

\end{enumerate}

\section*{Acknowledgements}
We acknowledge the anonymous referee for the detailed review and for the helpful suggestions, which allowed us to improve the manuscript. TCM acknowledges the FAPESP postdoctoral fellowship no. 2018/03480-7. TCM thanks Amanda Lopes for the help. APV acknowledges the FAPESP postdoctoral fellowship no. 2017/15893-1. We thank the kindly support to POLLY.







\clearpage
\twocolumn
\appendix
\section{Tables}\label{subsec:tables}

\subsection{Morphological Classification Tables}\label{subsubsec:morph_table}

\def\arraystretch{0.87}

\begin{table}
\centering
\caption{A brief resume of the possible classes  attributed in Galaxy Zoo 2. More details at \citet{Zoo2}.}
\label{tab:zooIIclass}
\begin{tabular}{cc}
\hline\hline
Class & Description \\
\hline
Ec   &  Elliptical completely round \\
Ec   &  Elliptical with cigar shape \\
Ei   &  Elliptical with shape between round and cigar \\
\hline
Ser  & Edge-on spiral with round bulge \\
Seb  & Edge-on spiral with boxy shape bulge \\
Sen  & Edge-on spiral with no bulge \\ 
\hline
Sa   & Spiral with dominant bulge \\
Sb   & Spiral with obvious bulge \\
Sc   & Spiral with noticeable bulge \\
Sd   & Spiral with no bulge \\
\hline
SBa  &  Barred spiral with dominant bulge \\
SBb  &  Barred spiral with obvious bulge \\
SBc  &  Barred spiral with just noticeable bulge \\
SBd  &  Barred spiral with no bulge \\
\hline
\hline
\end{tabular}
\end{table}

\clearpage

\onecolumn
\subsection{Dynamics Properties}{\label{Din_tab}}

    \begin{longtable}{lccccccccr}
    \caption{Dynamical parameters for the CGs of the sample. In each column we have: (1) the identification of each group; (2) number of members in each group; (3) total absolute magnitude of the group in the r-band; (4) velocity dispersion of the group; (5) harmonic radius (Mpc);  (6) total dynamical mass (in $M_{\odot}$); (7) the crossing time (in H$_{0}^{-1}$); (8) spatial  density; (9) spiral fraction and (10) dynamical class, either low-$\sigma$ (L) or high-$\sigma$ (H).}
    \label{tab:dynamics}
    \\
    \hline
    \multicolumn{1}{c}{GroupID} & \multicolumn{1}{c}{N${m}$} & \multicolumn{1}{c}{M$_{r,G}$} & \multicolumn{1}{c}{$\sigma$} & \multicolumn{1}{c}{R$_{harm}$} &
    \multicolumn{1}{c}{log M} & \multicolumn{1}{c}{$t_{c}$} & \multicolumn{1}{c}{log$\rho$} & \multicolumn{1}{c}{f$_{sp}$} & \multicolumn{1}{c}{Class}\\ \hline
    \endfirsthead
    
    \multicolumn{10}{c}%
{{\bfseries \tablename\ \thetable{} -- continued from previous page}} \\
\hline  \multicolumn{1}{c}{GroupID} & \multicolumn{1}{c}{N${m}$} & \multicolumn{1}{c}{M$_{r,G}$} & \multicolumn{1}{c}{$\sigma$} & \multicolumn{1}{c}{R$_{harm}$} &
    \multicolumn{1}{c}{log M} & \multicolumn{1}{c}{$t_{c}$} & \multicolumn{1}{c}{log$\rho$} & \multicolumn{1}{c}{f$_{sp}$} & \multicolumn{1}{c}{Class} \\ \hline 
\endhead

\hline 
\endfoot
\hline \hline
\endlastfoot

1327 & 4 & -21.32059 & 164.30531 & 0.03422 & 11.80918 & 0.01531 & 4.3773 & 0.5 & L \\
  70 & 4 & -23.32746 & 155.6021 & 0.0516 & 11.94032 & 0.02438 & 3.84206 & 0.25 & L \\
  510 & 4 & -22.23048 & 153.97421 & 0.0631 & 12.01857 & 0.03012 & 3.57991 & 0.25 & L \\
  326 & 4 & -23.43101 & 161.23244 & 0.06725 & 12.08627 & 0.03066 & 3.49682 & 0.0 & L \\
  820 & 4 & -22.66269 & 92.04479 & 0.03939 & 11.367 & 0.03146 & 4.19393 & 0.75 & L \\
  2209 & 4 & -22.17205 & 158.51787 & 0.07893 & 12.14104 & 0.0366 & 3.28828 & 0.75 & L \\
  321 & 4 & -23.09429 & 153.17699 & 0.08118 & 12.12346 & 0.03896 & 3.2517 & 0.0 & L \\
  389 & 4 & -22.52972 & 159.66516 & 0.09252 & 12.21629 & 0.04259 & 3.08133 & 0.0 & L \\
  113 & 4 & -22.46142 & 129.23766 & 0.09071 & 12.02408 & 0.05159 & 3.10703 & 0.5 & L \\
  633 & 4 & -23.07535 & 130.62134 & 0.10499 & 12.09681 & 0.05908 & 2.91659 & 0.25 & L \\
  1114 & 6 & -22.67863 & 158.65254 & 0.13108 & 12.36209 & 0.06074 & 2.80343 & 0.33 & L \\
  1616 & 4 & -22.49507 & 170.37962 & 0.14116 & 12.45621 & 0.0609 & 2.53081 & 0.25 & L \\
  252 & 4 & -23.55476 & 134.75757 & 0.12162 & 12.18777 & 0.06634 & 2.72494 & 0.25 & L \\
  904 & 4 & -23.23118 & 144.67075 & 0.13194 & 12.28478 & 0.06704 & 2.61886 & 0.25 & L \\
  1409 & 4 & -23.17326 & 121.24132 & 0.11638 & 12.07683 & 0.07056 & 2.78233 & 0.25 & L \\
  1063 & 4 & -22.63152 & 104.22949 & 0.10803 & 11.91319 & 0.07619 & 2.87931 & 0.75 & L \\
  1458 & 4 & -22.29223 & 146.67732 & 0.15218 & 12.35874 & 0.07627 & 2.43288 & 0.25 & L \\
  1385 & 4 & -22.77454 & 144.35689 & 0.15482 & 12.35237 & 0.07884 & 2.41046 & 0.0 & L \\
  1059 & 4 & -22.41693 & 134.353 & 0.1444 & 12.25973 & 0.07901 & 2.50123 & 0.0 & L \\
  724 & 4 & -23.37203 & 176.75452 & 0.19093 & 12.61926 & 0.07941 & 2.13736 & 0.0 & L \\
  1185 & 4 & -23.22581 & 153.71902 & 0.16608 & 12.43742 & 0.07942 & 2.31903 & 0.5 & L \\
  425 & 4 & -22.74902 & 165.96463 & 0.1797 & 12.53824 & 0.0796 & 2.21629 & 0.25 & L \\
  1011 & 4 & -23.62991 & 156.5828 & 0.17335 & 12.47206 & 0.08138 & 2.26323 & 0.25 & L \\
  728 & 4 & -21.8888 & 78.10072 & 0.08706 & 11.56879 & 0.08194 & 3.16049 & 0.0 & L \\
  2139 & 5 & -22.6674 & 171.11082 & 0.19777 & 12.60637 & 0.08496 & 2.1884 & 0.4 & L \\
  1090 & 4 & -22.38491 & 119.03671 & 0.14001 & 12.14117 & 0.08646 & 2.54151 & 0.25 & L \\
  1434 & 4 & -22.13416 & 128.16104 & 0.15158 & 12.23982 & 0.08695 & 2.438 & 0.25 & L \\
  1895 & 5 & -23.5568 & 108.89929 & 0.14542 & 12.08033 & 0.09816 & 2.58899 & 0.4 & L \\
  1767 & 5 & -23.67012 & 94.37662 & 0.13608 & 11.92716 & 0.10599 & 2.67553 & 0.2 & L \\
  952 & 4 & -23.06343 & 151.19118 & 0.22288 & 12.55078 & 0.10837 & 1.93575 & 0.5 & L \\
  1004 & 4 & -23.16966 & 172.05598 & 0.27225 & 12.74996 & 0.11632 & 1.67508 & 0.25 & L \\
  1539 & 4 & -22.97823 & 117.10299 & 0.19011 & 12.2598 & 0.11934 & 2.14293 & 0.5 & L \\
  1783 & 4 & -22.59076 & 57.29758 & 0.0973 & 11.34805 & 0.12484 & 3.0156 & 0.5 & L \\
  1020 & 4 & -22.65908 & 97.0623 & 0.17318 & 12.05625 & 0.13116 & 2.26447 & 0.0 & L \\
  2021 & 4 & -22.96518 & 156.2697 & 0.28007 & 12.67867 & 0.13175 & 1.63816 & 0.75 & L \\
  2011 & 4 & -23.70521 & 94.6297 & 0.17977 & 12.05043 & 0.13965 & 2.21582 & 0.5 & L \\
  1886 & 4 & -22.30497 & 119.47761 & 0.22944 & 12.35889 & 0.14116 & 1.89798 & 1.0 & L \\
  1163 & 4 & -23.14993 & 82.52488 & 0.16134 & 11.88458 & 0.14372 & 2.35671 & 0.75 & L \\
  1371 & 5 & -23.5258 & 124.59372 & 0.24885 & 12.43058 & 0.14682 & 1.88909 & 0.0 & L \\
  2155 & 4 & -21.92206 & 120.5286 & 0.24101 & 12.38787 & 0.14699 & 1.83386 & 0.0 & L \\
  1202 & 4 & -23.53411 & 80.5443 & 0.16772 & 11.8803 & 0.15307 & 2.30625 & 0.25 & L \\
  1553 & 4 & -22.93678 & 92.58204 & 0.19447 & 12.06557 & 0.15441 & 2.11338 & 0.5 & L \\
  1153 & 4 & -23.24351 & 102.1831 & 0.22063 & 12.20607 & 0.15872 & 1.949 & 0.5 & L \\
  1858 & 4 & -23.18516 & 59.49733 & 0.13608 & 11.52643 & 0.16813 & 2.57863 & 0.5 & L \\
  1109 & 4 & -22.67949 & 88.81782 & 0.21703 & 12.07718 & 0.17963 & 1.97039 & 1.0 & L \\
  1667 & 4 & -22.44853 & 97.20513 & 0.23881 & 12.19709 & 0.1806 & 1.84579 & 0.75 & L \\
  1605 & 4 & -22.83016 & 80.76373 & 0.20159 & 11.96256 & 0.18349 & 2.06655 & 0.25 & L \\
  1075 & 4 & -22.80477 & 91.26037 & 0.24006 & 12.14454 & 0.19337 & 1.83901 & 0.0 & L \\
  2149 & 4 & -22.63643 & 81.35825 & 0.22432 & 12.01533 & 0.20269 & 1.92734 & 0.0 & L \\
  2078 & 4 & -22.92845 & 107.98608 & 0.32292 & 12.41948 & 0.21982 & 1.4527 & 0.25 & L \\
  2176 & 4 & -23.13365 & 92.03265 & 0.2929 & 12.23825 & 0.23395 & 1.57982 & 0.0 & L \\
  1264 & 4 & -22.7586 & 39.24533 & 0.12627 & 11.13254 & 0.23652 & 2.67604 & 0.0 & L \\
  1213 & 5 & -23.26164 & 59.47038 & 0.19653 & 11.68568 & 0.24293 & 2.1966 & 0.2 & L \\
  1987 & 5 & -23.24915 & 66.6158 & 0.23337 & 11.85885 & 0.25752 & 1.97277 & 0.6 & L \\
  2202 & 4 & -23.80747 & 54.25206 & 0.42696 & 11.94288 & 0.57853 & 1.08879 & 0.0 & L \\
  1217 & 4 & -23.39339 & 579.8492 & 0.08246 & 13.28653 & 0.01045 & 3.23124 & 0.25 & H \\
  559 & 5 & -23.52758 & 578.21375 & 0.0878 & 13.31133 & 0.01116 & 3.24637 & 0.4 & H \\
  42 & 4 & -23.35465 & 429.73114 & 0.07248 & 12.97029 & 0.0124 & 3.39925 & 0.0 & H \\
  481 & 4 & -22.98172 & 264.16525 & 0.04732 & 12.36249 & 0.01317 & 3.9547 & 0.5 & H \\
  90 & 4 & -23.77042 & 417.0572 & 0.08009 & 12.9876 & 0.01412 & 3.26931 & 0.0 & H \\
  1265 & 4 & -21.96506 & 205.35538 & 0.04085 & 12.07988 & 0.01462 & 4.14632 & 0.25 & H \\
  1494 & 4 & -23.76876 & 559.589 & 0.11131 & 13.38593 & 0.01462 & 2.84035 & 0.5 & H \\
  841 & 4 & -22.76582 & 286.4665 & 0.06344 & 12.56013 & 0.01628 & 3.57297 & 0.0 & H \\
  353 & 4 & -23.07829 & 358.916 & 0.08182 & 12.8665 & 0.01676 & 3.24139 & 0.0 & H \\
  236 & 4 & -23.10858 & 234.0303 & 0.05883 & 12.35182 & 0.01848 & 3.6711 & 0.0 & H \\
  46 & 4 & -24.49025 & 241.84322 & 0.06581 & 12.42904 & 0.02001 & 3.525 & 0.0 & H \\
  177 & 4 & -22.68912 & 292.41367 & 0.08037 & 12.68072 & 0.0202 & 3.26476 & 0.5 & H \\
  1336 & 4 & -24.17166 & 518.2932 & 0.16047 & 13.4782 & 0.02276 & 2.36379 & 0.25 & H \\
  774 & 4 & -23.71035 & 210.6925 & 0.06763 & 12.3211 & 0.0236 & 3.48952 & 0.0 & H \\
  1036 & 4 & -22.01109 & 270.4543 & 0.08887 & 12.65661 & 0.02416 & 3.13365 & 0.0 & H \\
  711 & 4 & -22.586 & 301.06757 & 0.09974 & 12.79987 & 0.02435 & 2.9833 & 0.0 & H \\
  565 & 4 & -23.43296 & 347.75613 & 0.11626 & 12.99164 & 0.02458 & 2.78365 & 0.25 & H \\
  1249 & 4 & -23.48264 & 290.62234 & 0.09853 & 12.76389 & 0.02492 & 2.99924 & 0.25 & H \\
  2027 & 5 & -22.18198 & 467.63098 & 0.16563 & 13.40259 & 0.02604 & 2.41948 & 0.0 & H \\
  594 & 5 & -22.39405 & 279.22214 & 0.10583 & 12.76016 & 0.02786 & 3.00304 & 0.0 & H \\
  1407 & 4 & -23.86641 & 498.62305 & 0.19145 & 13.52126 & 0.02823 & 2.13378 & 0.0 & H \\
  508 & 4 & -22.76212 & 302.05862 & 0.11641 & 12.86983 & 0.02833 & 2.78199 & 0.0 & H \\
  1372 & 4 & -22.71358 & 326.48636 & 0.12942 & 12.98337 & 0.02914 & 2.644 & 0.75 & H \\
  2056 & 4 & -21.51397 & 247.17389 & 0.09902 & 12.62538 & 0.02945 & 2.99281 & 0.75 & H \\
  1065 & 4 & -22.43837 & 407.35153 & 0.16523 & 13.28167 & 0.02982 & 2.32573 & 1.0 & H \\
  933 & 4 & -22.30191 & 191.67117 & 0.07827 & 12.30235 & 0.03002 & 3.29922 & 0.75 & H \\
  2083 & 4 & -22.49336 & 340.31396 & 0.14233 & 13.06071 & 0.03074 & 2.52009 & 0.25 & H \\
  773 & 5 & -22.77582 & 303.24048 & 0.1353 & 12.93852 & 0.0328 & 2.68299 & 0.0 & H \\
  1390 & 5 & -23.84172 & 298.29538 & 0.13478 & 12.92256 & 0.03321 & 2.68804 & 0.2 & H \\
  1169 & 5 & -23.70944 & 192.53825 & 0.08835 & 12.35889 & 0.03373 & 3.23826 & 0.0 & H \\
  657 & 4 & -23.36415 & 301.94986 & 0.13968 & 12.94867 & 0.03401 & 2.54453 & 0.25 & H \\
  380 & 4 & -23.29121 & 284.13666 & 0.1324 & 12.8726 & 0.03425 & 2.61428 & 0.75 & H \\
  850 & 4 & -23.4875 & 442.3833 & 0.21142 & 13.46039 & 0.03513 & 2.00455 & 0.0 & H \\
  1713 & 4 & -22.87173 & 291.43262 & 0.13974 & 12.91804 & 0.03525 & 2.54403 & 0.5 & H \\
  1705 & 4 & -22.3627 & 275.7555 & 0.13341 & 12.84989 & 0.03556 & 2.60441 & 0.0 & H \\
  618 & 4 & -22.9252 & 226.38466 & 0.11216 & 12.60319 & 0.03642 & 2.83045 & 0.0 & H \\
  1214 & 4 & -23.99199 & 210.16231 & 0.1054 & 12.51161 & 0.03687 & 2.91143 & 0.25 & H \\
  596 & 4 & -23.20133 & 273.20392 & 0.13904 & 12.85976 & 0.03741 & 2.55056 & 0.25 & H \\
  209 & 4 & -22.26273 & 242.17967 & 0.12405 & 12.70553 & 0.03765 & 2.69917 & 0.25 & H \\
  225 & 4 & -23.32515 & 255.17973 & 0.13352 & 12.78288 & 0.03846 & 2.60335 & 0.0 & H \\
  375 & 4 & -22.17997 & 208.71983 & 0.11107 & 12.52836 & 0.03912 & 2.84323 & 0.5 & H \\
  1686 & 4 & -23.84638 & 245.40575 & 0.13142 & 12.74207 & 0.03937 & 2.62402 & 0.25 & H \\
  663 & 4 & -23.1078 & 303.56326 & 0.16704 & 13.03097 & 0.04045 & 2.3115 & 0.0 & H \\
  1303 & 4 & -22.29249 & 220.34158 & 0.12253 & 12.61809 & 0.04088 & 2.71522 & 0.25 & H \\
  2256 & 4 & -24.45498 & 251.71504 & 0.1405 & 12.79313 & 0.04103 & 2.53698 & 0.0 & H \\
  800 & 4 & -22.71132 & 232.23518 & 0.12979 & 12.68875 & 0.04108 & 2.64025 & 0.25 & H \\
  406 & 4 & -23.74362 & 240.54518 & 0.13825 & 12.74671 & 0.04225 & 2.55799 & 0.25 & H \\
  2087 & 4 & -23.05776 & 185.30817 & 0.11045 & 12.42261 & 0.04382 & 2.85046 & 0.0 & H \\
  253 & 4 & -23.08334 & 230.19572 & 0.14081 & 12.71649 & 0.04497 & 2.53403 & 0.75 & H \\
  673 & 4 & -23.57003 & 258.4045 & 0.16068 & 12.87422 & 0.04571 & 2.36207 & 0.0 & H \\
  135 & 4 & -23.01472 & 199.39705 & 0.12633 & 12.54459 & 0.04657 & 2.67546 & 0.25 & H \\
  2166 & 5 & -24.16096 & 469.9154 & 0.29866 & 13.66287 & 0.04672 & 1.65135 & 0.4 & H \\
  1717 & 4 & -23.24431 & 383.56552 & 0.24466 & 13.39989 & 0.04689 & 1.8143 & 0.25 & H \\
  1324 & 4 & -23.33909 & 260.70352 & 0.16742 & 12.89976 & 0.04721 & 2.30852 & 0.25 & H \\
  670 & 5 & -24.18419 & 377.83524 & 0.24745 & 13.39175 & 0.04814 & 1.89641 & 0.2 & H \\
  748 & 5 & -23.80492 & 318.838 & 0.21586 & 13.18497 & 0.04977 & 2.07436 & 0.2 & H \\
  895 & 4 & -22.32889 & 194.40796 & 0.14087 & 12.5699 & 0.05327 & 2.53351 & 0.0 & H \\
  1070 & 4 & -22.89126 & 248.3479 & 0.18542 & 12.90193 & 0.05488 & 2.17549 & 0.0 & H \\
  1274 & 4 & -22.9272 & 275.12704 & 0.20876 & 13.04236 & 0.05578 & 2.02105 & 0.5 & H \\
  1113 & 4 & -23.30823 & 218.56764 & 0.16728 & 12.74627 & 0.05626 & 2.30962 & 0.25 & H \\
  2208 & 5 & -23.81562 & 276.8277 & 0.21869 & 13.0679 & 0.05807 & 2.05741 & 0.2 & H \\
  1300 & 4 & -22.30241 & 183.40327 & 0.15063 & 12.54837 & 0.06037 & 2.44627 & 0.5 & H \\
  920 & 4 & -23.52712 & 286.63342 & 0.23863 & 13.13602 & 0.0612 & 1.84682 & 0.25 & H \\
  1592 & 4 & -22.70905 & 191.5309 & 0.16128 & 12.61572 & 0.0619 & 2.35719 & 0.5 & H \\
  1769 & 4 & -23.88926 & 283.71918 & 0.24039 & 13.13035 & 0.06228 & 1.83722 & 0.25 & H \\
  1464 & 4 & -22.71718 & 215.32292 & 0.18275 & 12.77169 & 0.06239 & 2.1944 & 0.5 & H \\
  811 & 4 & -22.4584 & 222.37167 & 0.18961 & 12.81566 & 0.06268 & 2.14642 & 0.0 & H \\
  1173 & 4 & -23.81617 & 276.80832 & 0.2416 & 13.1111 & 0.06416 & 1.83069 & 0.0 & H \\
  1909 & 5 & -23.94225 & 298.64487 & 0.26658 & 13.21979 & 0.06562 & 1.79941 & 0.0 & H \\
  1012 & 4 & -22.36551 & 210.87572 & 0.19765 & 12.78761 & 0.0689 & 2.09226 & 0.5 & H \\
  1301 & 4 & -23.52054 & 257.65 & 0.2492 & 13.06226 & 0.0711 & 1.79033 & 0.25 & H \\
  735 & 4 & -23.53957 & 211.2864 & 0.20515 & 12.80547 & 0.07138 & 2.04374 & 0.25 & H \\
  1189 & 5 & -23.13815 & 233.37009 & 0.22703 & 12.93583 & 0.07151 & 2.00861 & 0.8 & H \\
  1789 & 4 & -23.51063 & 287.02997 & 0.28038 & 13.20725 & 0.07181 & 1.63673 & 0.25 & H \\
  1901 & 4 & -23.27581 & 246.73026 & 0.24526 & 13.01772 & 0.07307 & 1.8111 & 0.25 & H \\
  954 & 4 & -22.93381 & 232.8592 & 0.23765 & 12.95378 & 0.07502 & 1.85215 & 0.75 & H \\
  1532 & 4 & -23.48681 & 290.3558 & 0.3017 & 13.24909 & 0.07638 & 1.54125 & 0.25 & H \\
  382 & 4 & -23.66696 & 235.03648 & 0.25684 & 12.99559 & 0.08033 & 1.75098 & 0.0 & H \\
  1551 & 4 & -23.05234 & 218.18669 & 0.23852 & 12.89883 & 0.08036 & 1.84741 & 0.25 & H \\
  2225 & 5 & -23.31458 & 271.63577 & 0.29852 & 13.1866 & 0.08079 & 1.65197 & 0.4 & H \\
  2295 & 4 & -23.13318 & 379.60504 & 0.41893 & 13.62446 & 0.08113 & 1.11354 & 0.25 & H \\
  1487 & 4 & -23.21317 & 236.5384 & 0.26164 & 13.00915 & 0.08131 & 1.72688 & 0.25 & H \\
  2263 & 7 & -24.03493 & 309.5383 & 0.34409 & 13.36175 & 0.08172 & 1.61299 & 0.43 & H \\
  2067 & 5 & -23.32188 & 187.8088 & 0.22045 & 12.73439 & 0.08629 & 2.04698 & 0.2 & H \\
  1779 & 4 & -24.19759 & 349.39413 & 0.44331 & 13.57699 & 0.09327 & 1.03984 & 0.25 & H \\
  737 & 4 & -23.46191 & 182.11754 & 0.23148 & 12.72887 & 0.09344 & 1.88643 & 0.25 & H \\
  1764 & 4 & -23.34622 & 186.03825 & 0.24803 & 12.77735 & 0.098 & 1.79648 & 0.0 & H \\
  533 & 4 & -23.43282 & 205.0583 & 0.27449 & 12.90593 & 0.0984 & 1.66439 & 0.25 & H \\
  2284 & 4 & -22.62204 & 182.58755 & 0.25591 & 12.77468 & 0.10303 & 1.75571 & 0.0 & H \\
  1341 & 4 & -24.649 & 227.4405 & 0.32053 & 13.06326 & 0.1036 & 1.46235 & 0.0 & H \\
  2102 & 4 & -24.14641 & 192.48509 & 0.27218 & 12.8473 & 0.10395 & 1.67541 & 0.0 & H \\
  2099 & 4 & -23.58294 & 249.14816 & 0.35999 & 13.19286 & 0.10622 & 1.31109 & 0.0 & H \\
  1044 & 4 & -23.70229 & 205.87936 & 0.30124 & 12.9498 & 0.10756 & 1.54321 & 0.25 & H \\
  1681 & 4 & -23.29212 & 184.97379 & 0.32199 & 12.88571 & 0.12796 & 1.45646 & 0.0 & H \\
  1388 & 4 & -23.85669 & 197.21857 & 0.38342 & 13.01723 & 0.14292 & 1.22893 & 0.5 & H \\
  2239 & 4 & -23.62445 & 217.1688 & 0.44853 & 13.16904 & 0.15182 & 1.02461 & 0.25 & H \\
    \end{longtable}
    



\bsp	
\label{lastpage}
\end{document}